\documentclass[nohyper]{JHEP3}
\usepackage{amssymb}

\newcommand{\bmat}{\left(\begin{array}}
\newcommand{\emat}{\end{array}\right)}

\def\yzero{\smash{\hbox{$y\kern-4pt\raise1pt\hbox{${}^\circ$}$}}}

\def\-{\hphantom{-}}

\def\beq{\begin{equation}}
\def\eeq{\end{equation}}
\def\ba{\begin{eqnarray}}
\def\ea{\end{eqnarray}}
\def\s2{\frac{1}{2}}

\def\beq{\begin{equation}}
\def\eeq{\end{equation}}
\def\beqa{\begin{eqnarray}}
\def\eeqa{\end{eqnarray}}

\def\IF{\relax{\rm I\kern-.18em F}}
\def\II{\relax{\rm I\kern-.18em I}}
\def\IP{\relax{\rm I\kern-.18em P}}
\def\IC{\relax\hbox{\kern.25em$\inbar\kern-.3em{\rm C}$}}
\def\IR{\relax{\rm I\kern-.18em R}}

\def\Dsl{\,\raise.15ex\hbox{/}\mkern-13.5mu D} 

\def\IC{\bf C}

\title {Winding Strings in AdS$_{3}$}

\author {Estanislao Herscovich$^{a,d}$,
Pablo 
Minces$^{a,b}$ and Carmen N\'u\~nez$^{a,c}$\\
$^{a}$Instituto de Astronom\'{\i}a y F\'{\i}sica del Espacio 
(IAFE),\\
C.C.67 - Suc. 28, 1428 Buenos Aires, Argentina.\\
$^{b}$Centro Brasileiro de Pesquisas F\'{\i}sicas (CBPF),\\
Departamento de Teoria de Campos e Part\'{\i}culas 
(DCP),\\Rua Dr. Xavier Sigaud 150, 22290-180, Rio de Janeiro, RJ, 
Brazil.\\
$^{c}$Physics and $^d$Mathematics Department, University of Buenos 
Aires,\\
Ciudad Universitaria, Pab. I, 1428 Buenos Aires, Argentina}

\abstract{Correlation functions of one-unit spectral flowed states in
string theory on AdS$_3$ are considered. We present
the modified Knizhnik-Zamolodchikov and null 
vector equations to be satisfied by amplitudes containing  states in 
winding sector one and study their solution corresponding to 
the four point function including one $w=1$ field.
We compute the three point function involving two
one-unit spectral flowed operators and find
expressions for amplitudes  of three 
$w=1$ states satisfying certain 
particular relations among the spins of the fields. Several consistency 
checks are performed.}

\begin{document}

\section{Introduction}
Correlation functions of operators creating string states in 
three dimensional anti-de Sitter space (AdS$_3$) are 
essential ingredients to establish the consistency of string theory in this
geometry as well as to explore the AdS/CFT correspondence beyond the 
supergravity approximation. However 
 the non-rational structure of the SL(2,R) CFT describing the worldsheet of 
strings propagating in AdS$_3$ presents some difficulties to
the computation of these correlators, 
since very little is known about non-compact conformal field theories in 
general.

Nevertheless, amplitudes of some physical states 
were computed in reference \cite{malda3}. 
The starting point in these calculations is the SL(2,C)/SU(2) WZW model 
which describes the worldsheet of strings 
propagating in the hyperbolic space H$_3$. 
Two, three and four point
functions of primary fields in this coset model were derived in references 
\cite{tesch1, tesch2}. The connection between these correlators and some 
amplitudes in
 the Lorentzian theory described by the
SL(2,R) WZW model was performed in \cite{malda3} using the equivalence 
between
string theory on AdS$_3$ and the dual two dimensional CFT on the boundary.
Actually, the interpretation of these amplitudes
as correlation functions of the dual CFT is crucial to determine the
correlators and to establish
the structure of the factorization of  four point functions. 
In this way the
closure of the operator algebra on the Hilbert space of string theory on 
AdS$_3$ \cite{malda1,malda2} was verified in \cite{malda3} for four point
functions of  primary fields related to primaries of the SL(2,C)/SU(2) 
coset model through analytic continuation.

However not all the states in the physical spectrum of string theory on 
AdS$_3$ can be obtained analytically continuing fields in H$_3$. Actually,
the spectral flow symmetry of the SL(2,R) WZW model establishes the 
occurrence of winding states 
created by spectral flowed operators \cite{malda1}. 
An important auxiliary 
tool to construct these states is the spectral flow operator which allows 
to interpolate
between objects in different winding sectors $w$. 
Two and certain three point
 functions containing spectral flowed
 fields were computed in \cite{malda3} and used 
to
verify both the factorization of four point functions of $w=0$ fields 
into products
of three point functions summed over intermediate physical states and the
pattern of spectral flow number conservation of the
 correlators determined by the SL(2,R) current algebra.

Two alternative methods were proposed in \cite{malda3}
to evaluate $N$-point functions containing 
states in winding sectors $w\ne 0$.
Both procedures  involve the insertion of 
one auxiliary spectral flow operator for each winding unit.
This implies
the calculation of expectation values of more than $N$ vertex operators.
 The two approaches were applied to 
compute two point functions in arbitrary winding sectors and
three point functions involving one $w=1$ 
 and two $w=0$ states, but more general three and four point functions
 require the calculation of correlators with five or more operator insertions,
with the consequent complications.
These general 
amplitudes are needed to definitely settle the question about the 
unitarity of the theory. Indeed the structure of the factorization of 
four point functions should be consistent with the physical Hilbert space 
of string theory, but four point functions involving spectral flowed 
states have not been computed so far. In this paper we give one step 
forward towards 
this project by computing the three point function 
containing two $w=1$ states from the five point function
involving two spectral flow operators. We also study the four point function 
including one $w=1$ state starting from the five point function 
containing only one spectral flow operator. Besides we find
expressions for amplitudes  
 of three $w=1$ states satisfying certain particular relations among the 
spins of the fields. 

As is well known expectation values of fields in WZW models
must obey the Knizhnik-Zamolodchikov (KZ) 
equations \cite{zamo2}, a system of linear differential
equations which follow from the Sugawara
construction of the energy-momentum tensor. 
An important additional
property of WZW models for compact groups 
is the existence of null vectors in the
Verma modules of the primary fields. These give additional differential
equations which allow to determine the fusion rules and eventually solve
the theory \cite{zamo1}. Unfortunately the unitary representations of 
SL(2,R)
which give rise to the physical spectrum of string theory on AdS$_3$
do not contain singular vectors. Nevertheless the spectral flow operator 
has a null current algebra descendant which plays a relevant role 
 in this non-rational theory. Indeed it adds one  differential 
equation for each unit of spectral flow of the operators involved in the 
amplitudes. Moreover this null state allows to simplify the KZ equations
in the coordinates labeling the position of the spectral flow operators
\cite{malda3}.

Despite all this information the differential equations to be obeyed by
correlation functions containing spectral flowed operators are difficult to
solve because, as we will see, they turn out to give iterative relations. 
Actually we shall show in the following sections that
the KZ and null vector equations for amplitudes of
states in winding sectors $w\ne 0$
relate two or more expectation values in which the spectral
flowed fields at a given position
 have different spins and conformal dimensions.  
Nonetheless we shall 
 manipulate and solve the system of iterative equations 
in certain particular cases needed
to obtain correlators of three  $w=1$ 
string states and four point functions involving one $w=1$ state.

The organization of this paper is as follows.
For completeness and in order to introduce our notations and conventions,
in the next section we briefly review the spectrum 
of string theory on 
AdS$_3$, recall the construction of one-unit spectral flowed 
operators and collect the results of the 
amplitudes involving such fields which have been obtained
so far.
In section 3 we compute the three point function involving two $w=1$ 
fields and study its pole structure. 
As a consistency check, we verify that it correctly reproduces the 
two point function of a $w=1$ field when the third operator is the 
identity. In addition we find, as a byproduct, the four point function 
including one $w=1$ field along with a spectral flow operator.
In section 4, we discuss the 
Ward identities to be satisfied by correlation
functions containing $w=1$ fields and we deduce the modified KZ and null 
vector equations that they must obey. The solution to these equations
is analyzed in section 5 for the case of the
four point function of one
$w=1$ field and three unflowed generic states. 
This is done by first expanding the correlator in powers of 
the corresponding cross ratio coordinate. We explicitly find the lowest 
order contribution and write an iterative differential equation for the higher 
orders in terms of the lowest one. Two consistency checks are 
succesfully performed. First 
we verify that, for appropriate choices of the spins, 
the correlator properly reduces to the three point 
function involving one $w=1$ field, as computed in \cite{malda3}. Then we also
verify that the functional form of the
four point function including one $w=1$ field and  a spectral flow 
operator computed in section 3 is correctly reproduced.
Conclusions and discussions are offered in Section 6. 

We have included three appendices.
Some properties of 
five and six point functions 
containing states with generic spin along with spectral flow 
operators
are listed in Appendix A, whereas Appendix B presents some useful 
formulae which are used in the main body of the article. In Appendix C we 
compute 
expressions of the three point function involving three $w=1$ 
operators for certain particular relations among the spins of the 
fields. This is done by proposing an  {\it ansatz} 
for the solution of the modified KZ and null vector equations.

\section{Perturbative string theory on AdS$_3$}

In this section we gather known results about
 the spectrum  and correlation functions of
perturbative string theory on AdS$_3$ 
in order to set up our conventions. We follow the same notation as 
reference \cite{malda3} and so the expert reader can proceed directly to 
the next section.

\subsection{Notation and conventions}

The Hilbert space of the WZW model is a sum of products of representations 
of the SL(2,R) current algebra given by
\ba 
&&[J^{3}_{n},J^{3}_{m}]=-\frac{k}{2}\; n\delta_{n+m,0}\; ,\nonumber\\
&&[J^{3}_{n},J^{\pm}_{m}]=\pm J_{n+m}^{\pm}\; ,\nonumber\\
&&[J^{+}_{n},J^{-}_{m}]=-2J^{3}_{n+m}+kn\delta_{n+m,0}\; , \nonumber
\ea
and the same for $\bar J^{3,\pm}$. 
The Sugawara construction of the energy-momentum tensor
\ba
T(z)=\frac{1}{k-2}\; (J^{+}(z)J^{-}(z)-J^{3}(z)J^{3}(z))\; , 
\nonumber
\ea
determines the Virasoro generators
\ba
L_{0}&=&\frac{1}{k-2}\left[ 
\frac{1}{2}(J_{0}^{+}J_{0}^{-}+J_{0}^{-}J_{0}^{+})-(J_{0}^{3})^{2}
+\sum_{m=1}^{\infty}(J_{-m}^{+}J_{m}^{-}+J_{-m}^{-}J_{m}^{+}-
2J_{-m}^{3}J_{m}^{3})\right]\; , \nonumber\\
L_{n\not= 
0}&=&\frac{1}{k-2}\sum_{m=1}^{\infty}(J_{n-m}^{+}J_{m}^{-}+
J_{n-m}^{-}J_{m}^{+}-2J_{n-m}^{3}J_{m}^{3})\; , \nonumber
\ea
which obey the following commutation relations
\ba
[L_{n},L_{m}]=(n-m)L_{n+m}+\frac{c}{12}n(n^{2}-1)\delta_{n+m,0}\; . 
\nonumber
\ea
The central charge is given by
\ba
c=\frac{3k}{k-2}\; , \nonumber
\ea
with $k>2$.

The physical states of string theory on AdS$_3$ are in 
unitary representations of the universal cover of SL(2,R) \cite{malda1}. 
These are generated from 
the continuous (${\cal 
C}_j^\alpha$) 
and the lowest 
(${\cal D}^+_j$)
 and highest (${\cal D}^-_j$) weight discrete 
representations 
of 
the zero modes acting with $J^{3,\pm}_{-n}, n > 0$. In our conventions 
these are
\ba
{\cal D}_{j}^{+}&=&\{|j;m\rangle\;:\; m=j,\; j+1,\; j+2,\cdots\}\; ,
\nonumber\\
{\cal D}_{j}^{-}&=&\{|j;m\rangle\;:\; m=-j,\; -j-1,\; -j-2,\cdots\}\; ,
\nonumber
\ea
with 
\beq
\frac{1}{2}< j< \frac{k-1}{2}\; ,
\label{51}
\eeq
and
\ba
{\cal C}_{j}^{\alpha}=\{|j,\alpha ; m\rangle\;:\; m=\alpha,\; \alpha\pm 
1,\; 
\alpha\pm 2,\cdots , \alpha \in R \} \; , \qquad
j = \frac 12 + is \;, \quad s\in {\bf R} \; .
\nonumber
\ea

The conformal weight of the primary fields is given by
\beq
\Delta_j = -\frac{j(j-1)}{k-2}\; ,
\label{wo}
\eeq
and similarly for $\bar \Delta_{\bar j}$. The current algebra
descendants of the primary operators contribute an additional integer
for each excitation level.
In the string theory application one can consider the spacetime to be a 
product of AdS$_3$ times an internal manifold. In this case the conformal
weight of the physical states may be supplemented with a contribution
from the internal CFT, usually denoted $h$. Moreover 
the physical state conditions for string states, namely 
$L_0|\Psi>=|\Psi>,
 L_n|\Psi>=0, n>0$ and $L_0=\bar L_0$ determine $j=\bar j$. 

The spectral flow automorphism
\ba
{\tilde J}^{3}_{n}=J^{3}_{n}-\frac{k}{2}\;w\delta_{n,0}\; ,\qquad
{\tilde J}^{\pm}_{n}=J^{\pm}_{n\pm w}\; ,\qquad
{\tilde L}_{n}=L_{n}+w 
J^{3}_{n}-\frac{k}{4}\;w^{2}\delta_{n,0}\; ,
\nonumber
\ea
parametrized with $w\in {\bf Z}$, generates new representations defined by
\ba
&& J^{\pm}_{n\pm w}|j, m, w\rangle=0\; ,\qquad 
J^{3}_{n}|j, m, w\rangle=0\; ,\qquad (n\geq 1)\; ,\nonumber\\
&& J^{3}_{0}|j, m, w\rangle=\left(m+\frac{k}{2}\;w\right)| j, m, w \rangle\; .
\nonumber
\ea
The conformal 
weight of the  primary fields (\ref{wo}) transforms consequently as
\beq
\Delta^w_{j}=-\frac{ j( j-1)}{k-2}- m w 
-\frac{k}{4}\;w^{2}\; ,
\label{520}
\eeq
and similarly for $\bar\Delta_j^{\bar w}$.
Periodicity of
the closed string under the worldsheet coordinate transformation 
$\sigma\rightarrow \sigma+2\pi$ settles $w=\bar w$. 
These spectral flowed states have to be added to the
unflowed fields of the full representations generated from 
${\cal D}^\pm_j$ and ${\cal C}^\alpha _j$ in order to 
describe the complete spectrum of the theory.

The primary states in the sector $w=0$ can be represented by an operator 
$\Phi_j(x,\bar x;w,\bar w)$ which satisfies the following OPE with the 
currents
\ba
J^{a}(z)\Phi_{j}(x,{\bar x};w,{\bar w})\sim 
\frac{D^{a}}{z-w}\;\Phi_{j}(x,{\bar x};w,{\bar w})\; ,\qquad (a=3,\pm)\; ,
\nonumber
\ea
where the differential operators
\ba
D^{+}=\frac{\partial}{\partial x}\; ,\qquad 
D^{3}=x\frac{\partial}{\partial 
x}\; +\; j\; ,\qquad D^{-}=x^{2}\frac{\partial}{\partial
x}\; +\; 2jx\; ,
\nonumber
\ea
give a representation of the Lie algebra of SL(2).
Here $x, \bar x$ keep track of the SL(2) weights of the fields and 
they are interpreted as the coordinates of  
the boundary in the AdS/CFT context. 

One can also consider 
operators in the $m$ basis, obtained through the following 
transformation from the $x$ basis
\beq
\Phi_{j; m, \bar m} = \int \frac{d^2 x}{|x|^2}x^{j-m}\bar x^{j 
- \bar m} \Phi_{j}(x,\bar x) \; ,
\label{max}
\eeq
where $m-\bar m$ is an integer.

In the sector $w=1$ the spectral flowed states are constructed by 
the fusion of $\Phi_j$ with the spectral flow operator 
$\Phi_{\frac k2}$ through the following operation \cite{malda3}
\ba
&&\Phi^{w =1,j}_{J,\bar J}(x,\bar x; z, \bar z) \equiv 
\lim_{\epsilon,\bar \epsilon\rightarrow 0}
\epsilon^{m}\bar\epsilon^{{\bar m}}
\int d^{2}y\;
y^{j- m -1} \bar y^{j-{\bar m}-1} \nonumber\\
&&\qquad\qquad\qquad\qquad\quad\qquad \times \quad \Phi_{j}(x+y,\bar x + 
\bar y ; 
z+\epsilon, 
\bar z + \bar 
\epsilon)\Phi_{\frac k2}(x,\bar x ; z, \bar z)\; ,
\label{1}
\ea
where
\beq
J=m+\frac k2\; ,\qquad\qquad \bar J = \bar m + \frac k2\; ,
\label{1aa}
\eeq 
denote the left and right spins of the $w=1$ field. 
In 
the $x$ basis, the winding number
 $w$ turns out to be always positive, unlike  in 
the $m$ basis where the sign of $w$ is correlated with the sign of $m$, thus 
distinguishing by convention 
incoming from outgoing spectral flowed states in the correlation functions. 

Alternatively $\Phi_{J,\bar J}^{w=1, j}$ can be also defined by integrating 
over $\epsilon$ as
\ba
&&\Phi^{w =1,j}_{J,\bar J}(x,\bar x; z, \bar z) \equiv
\lim_{y,\bar y\rightarrow 0}
y^{j-m} \bar y^{j-\bar m}
\int d^{2}\epsilon\;
\epsilon^{m-1}
\bar \epsilon^{\bar m-1} \times \nonumber\\
&&\qquad\qquad\qquad\qquad\qquad \times\quad \Phi_{j}(x+y,\bar x + 
\bar y ; z+\epsilon, \bar z + 
\bar\epsilon)\Phi_{\frac{k}{2}}(x,\bar x; z, \bar z)\; .
\label{1a}
\ea
Both definitions (\ref{1}) and (\ref{1a}) are equivalent and they are 
understood 
to hold inside 
correlation functions. 
The first one is local in $z, \bar z$ 
whereas the 
second one is local in $x, \bar x$.
As 
pointed out in \cite{malda3}, the limit $\epsilon\rightarrow 0$ in 
(\ref{1}) $-$ and 
similarly $y\rightarrow 0$ in (\ref{1a}) $-$ exists 
and it verifies  several
important checks. For instance 
it has the right operator product expansions with 
the currents, namely
\ba
J(x^\prime , z^\prime )\Phi_{J,\bar J}^{w=1, j}(x,z) &=& -(j-m-1) 
\frac{(x-x^\prime )^2}{(z^\prime -z)^2}\; \Phi_{J+1, \bar J}^{w=1, j} 
(x,z) 
 \nonumber\\ &+&
\frac 1{z^\prime -z}\left [ (x-x^\prime )^2 \frac 
{\partial}{\partial x} + 2\left ( m + \frac k2 \right )(x-x^\prime )  
\right ] \Phi_{J,\bar J}^{w=1, j}(x,z) \; ,\nonumber\\
\label{current}
\ea
where
\ba
J(x,z)=-J^{-}(z)+2xJ^{3}(z)-x^{2}J^{+}(z)\; .
\nonumber
\ea
Furthermore the definition (\ref{1}) 
reproduces the 
expressions derived in 
\cite{malda3} for the two point function of spectral flowed states
and the three point function which includes in 
addition two other operators in the sector $w=0$. 

Vertex operators for string states in higher winding sectors can be
easily obtained in the $m$ basis where they are expressed in terms
of SL(2) parafermions and one free boson \cite{malda1} (see \cite{nunez1} 
for the free
field representation). However, as the winding number increases,
 they become more complicated in the $x$ basis.

\subsection{Correlation functions of winding states}

Correlation functions of unflowed states were computed in \cite{malda3} 
performing analytic continuation on the results for the Euclidean 
SL(2,C)/SU(2) WZW model obtained in \cite{tesch1, tesch2}. As discussed
 in \cite{malda3}, correlators 
including spectral flowed states can be evaluated in the $m$ basis 
starting from expectation values of states in the $w=0$ sector
including spectral flow operators. 
Alternatively one can perform the spectral flow 
operation directly in the $x$ basis using the definition (\ref{1}).

The two point function of 
spectral flowed states was computed in 
\cite{malda3}(see also \cite{nunez1} for a derivation in the 
$m$ basis using the free field theory approach)
and it is the following

\ba
&&\langle
\Phi_{J,\bar J}^{w,j}(x_1,z_1)
\Phi_{J,\bar J}^{w,j^\prime}(x_2,z_2)\rangle = 
x_{12}^{-2J}\bar x_{12}^{-2\bar J}
z_{12}^{-2\Delta^{w}_{j}}
\bar z_{12}^{-2{\bar\Delta}^{w}_{j}} \nonumber\\
&&\qquad\qquad \times \left [\delta(j+j^\prime -1) + \delta 
(j-j^\prime )\frac 
{\pi B(j)}{\gamma(2j)} \frac{\Gamma(j+m)}{\Gamma(1-j+m)} 
\frac{\Gamma(j-\bar m)}{\Gamma(1-j-\bar m)} \right ]\; , 
\label{op}
\ea
where $\Delta^{w}_{j}$ is given in (\ref{520}) and

\beq
B(j)=
\frac{k-2}{\pi}\frac{\nu^{1-2j}}{\gamma\left(\frac{2j-1}{k-2}\right)}\; 
,\quad 
\nu 
=\pi\;\frac{\Gamma\left(\frac{k-3}{k-2}\right)}
{\Gamma\left(\frac{k-1}{k-2}\right)}\; 
, \label{b2p}
\eeq

\beq
\gamma(x)\equiv\frac{\Gamma(x)}{\Gamma(1-x)}\; .
\label{370}
\eeq

Recall that in the $x$ basis the operators are labeled with positive $w$, 
so this two point function conserves winding number as expected. As is well 
known an $N$ point function can violate 
winding number by $N-2$ units 
\cite{malda3, zamo3}.\footnote{See Appendix D of \cite{malda3} for a 
detailed analysis.} 
The three point function 
including one operator in the $w =1$ sector is the following 
\footnote{Actually 
this expression differs from the one in \cite{malda3} by an irrelevant
factor $(-1)^{J-{\bar J}}$, as it can be verified using the property 
$J-{\bar J}\in {\bf Z}$ together with the identity 
$\Gamma(x)\Gamma(1-x)=\frac{\pi}{sin (\pi x)}$.} 
\ba
&&\left<\Phi^{w
=1,j_{1}}_{J,{\bar J}}(x_{1},z_{1})\Phi_{j_{2}}(x_{2},z_{2})
\Phi_{j_{3}}(x_{3},z_{3})\right> = \nonumber\\
&&\qquad \qquad = ~~
B(j_{1})
C\left(\frac{k}{2}-j_{1},j_{2},j_{3}\right)
\pi\;\frac{1}
{\gamma(j_{1}+j_{2}+j_{3}-k/2)}
\nonumber\\ &&\qquad\qquad \times ~~
\frac{\Gamma(j_{1}+J-\frac{k}{2})}{\Gamma(1+J-j_{2}-j_{3})}\;
\frac{\Gamma(j_{2}+j_{3}-{\bar J})}{\Gamma(1-j_{1}-{\bar
J}+\frac{k}{2})}
\nonumber\\ &&\qquad \qquad \times ~~
\left(x_{21}^{j_{3}-j_{2}-J}x_{31}^{j_{2}-j_{3}-J}
x_{32}^{J-j_{2}-j_{3}}\right)\left(
z_{21}^{\Delta_{3}-\Delta_{2}-\Delta^{w=1}_{1}}
z_{31}^{\Delta_{2}-\Delta_{3}-\Delta^{w=1}_{1}}
z_{32}^{\Delta^{w=1}_{1}-\Delta_{2}-\Delta_{3}}\right)
\nonumber\\ &&\qquad\qquad \times ~~ ({\rm antiholomorphic\; part})
\; ,
\label{3w}
\ea
where $J$ is given in (\ref{1aa}) and $\Delta^{w=1}_{1}$ is (see 
(\ref{520}))
\beq
\Delta^{w=1}_{1}=\Delta_{1}-J+\frac{k}{4}\; .
\label{3ww}
\eeq
$B(j)$ is given in (\ref{b2p}) 
and $C(j_{1},j_{2},j_{3})$ is the coefficient 
corresponding to the amplitude of three $w=0$ fields, namely 
\beq
C(j_{1},j_{2},j_{3})=-\frac{G(1-j_{1}-j_{2}-j_{3})G(j_{3}-j_{1}-j_{2})
G(j_{2}-j_{3}-j_{1})G(j_{1}-j_{2}-j_{3})}
{2\pi^{2}\nu^{j_{1}+j_{2}+j_{3}-1}
\gamma\left(\frac{k-1}{k-2}\right)G(-1)G(1-2j_{1})G(1-2j_{2})G(1-2j_{3})}\; 
,
\label{60}
\eeq
where 
\ba
G(j)=(k-2)^{\frac{j(k-1-j)}{2(k-2)}}\Gamma_{2}(-j\mid 
1,k-2)\Gamma_{2}(k-1+j\mid 1,k-2)\; ,
\nonumber
\ea
and $\Gamma_{2}(x|1,\omega)$ is the Barnes double Gamma function 
which reads
\ba
\log (\Gamma_{2}(x\mid 1,\omega))=\lim_{\epsilon\rightarrow 
0}\frac{\partial}{\partial\epsilon}\left[\sum_{n,m=0}^{\infty}
(x+n+m\omega)^{-\epsilon}-\sum_{n,m=0\; ;\; (n,m)\not= (0,0)}^{\infty}
(n+m\omega)^{-\epsilon}
\right]\; . 
\nonumber
\ea

This three point function
was obtained in \cite{malda3} by first computing a four 
point function including one spectral flow operator $\Phi_{\frac k2}$. 
Such  calculation is 
performed by explicitly solving the corresponding KZ and null vector 
equations. The four point function gives rise to (\ref{3w}) 
after spectral flowing as in the 
definition (\ref{1}) or alternatively, after transforming to the $m$ basis,
extracting the pole residue at $m=-\frac k2$ and acting with 
the spectral flow operator on the unflowed field $\Phi_{j_1}$. 

This summarizes the already known explicit expressions for correlators 
including spectral flowed fields. Note that, whereas the two point 
function is known for fields in unlimited winding sectors, the situation 
gets more complicated in the case of the three point function, where only 
the case involving one $w=1$ and two $w=0$ operators has been computed so 
far. The increasing difficulties to compute three point 
functions including additional spectral flowed fields are due to the 
fact that one has to start from amplitudes containing more 
spectral flow operators. 
In the following section we shall illustrate this statement by going 
one step further and computing the amplitude of two $w=1$ and
one $w=0$ states in the $x$ basis starting from the five point function
which includes two spectral flow operators. 

\section{Three point function involving two $w=1$ fields}

We want to compute the following three point function including two 
$w=1$ fields

\beq
\left<\Phi^{w =1,j_{1}}_{J_{1},{\bar J}_{1}}(x_{1},z_{1})
\Phi_{j_{2}}(x_{2},z_{2})
\Phi_{J_{3},{\bar J}_{3}}^{w=1, 
j_3}(x_{3},z_{3})\right> \; .
\label{tre2}
\eeq

The starting point is the five point function with two 
spectral flow operators, namely

\beq
A_{5}\equiv\langle 
\Phi_{\frac k2}(x_{1},z_{1})
\Phi_{\frac k2}(x_{2},z_{2})
\Phi_{j_{1}}(y_{1},\zeta_{1})
\Phi_{j_{2}}(y_{2},\zeta_{2})
\Phi_{j_{3}}(y_{3},\zeta_{3})
\rangle\; 
. \label{14}
\eeq
Due to the spectral flow operators inserted at 
$(x_{1},z_{1})$ 
and $(x_{2},z_{2})$, $A_5$ must obey the null vector equations
\beq
0=\sum_{i=1,\; i\not= 
1}^{5}\frac{x_{i}-x_{1}}{z_{1}-z_{i}}\left[(x_{i}-x_{1})
\frac{\partial}{\partial x_{i}}+2j_{i}\right]\; A_{5}\; ,
\label{16}
\eeq
at $(x_{1},z_{1})$ and a similar one 
at $(x_{2},z_{2})$, where we have renamed the insertion points
$y_i=x_{i+2}, \zeta_i=z_{i+2}$.
Moreover the singular state condition allows to 
simplify the action of the
Sugawara construction on $\Phi_{\frac k2}$ as 
\beq
L_{-1}|j=k/2\rangle=-J^{3}_{-1}|j=k/2\rangle .
\eeq
 Therefore the 
insertion of the spectral flow operators  
implies the following reduced form
of the corresponding KZ equation

\beq
\frac{\partial A_{5}}{\partial z_{1}} 
=-\sum_{i=1,\; i\not= 1}^{5}\frac{1}{z_{1}-z_{i}}\left[(x_{i}-x_{1})
\frac{\partial}{\partial x_{i}}+j_{i}\right]\; A_{5}\; ,
\label{15}
\eeq
and similarly for $(x_{2},z_{2})$. 

The $x_i$ dependence of 
the solution to the null vector equations
 was found in \cite{zamo3} (see also \cite{malda3}). 
Here we give the complete solution including
the dependence on the worldsheet coordinates 
$z_i$,
 which is determined from the Ward identities and the reduced KZ equations
(\ref{15}), namely

\ba
A_5&=&
B(j_1)B(j_3) C \left ( \frac k2 - j_1, j_2,
\frac k2 - j_3\right )
|z_{12}|^k |z_{13}|^{-2j_1}|z_{14}|^{-2j_2}|z_{15}|^{-2j_3}\nonumber\\
&&\quad\times |z_{23}|^{-2j_1}|z_{24}|^{-2j_2}|z_{25}|^{-2j_3}
|z_{34}|^{2(\Delta_3-\Delta_1-\Delta_2)}
|z_{35}|^{2(\Delta_2-\Delta_1-\Delta_3)}
|z_{45}|^{2(\Delta_1-\Delta_2-\Delta_3)}\nonumber\\
&&\quad \times |x_{12}|^{2(j_1+j_2+j_3-k)}
|\mu_1|^{2(j_1-j_2-j_3)}
|\mu_2|^{2(j_2-j_1-j_3)}|\mu_3|^{2(j_3-j_1-j_2)} \quad ,
\label{5pt}
\ea
with
\ba
\mu_1=\frac{x_{14}x_{25}}{z_{14}z_{25}}-\frac{x_{15}x_{24}}{z_{15}z_{24}} 
\quad ,
\nonumber\\
\mu_2=\frac{x_{15}x_{23}}{z_{15}z_{23}}-\frac{x_{13}x_{25}}{z_{13}z_{25}}
\quad ,
\nonumber\\
\mu_3=\frac{x_{13}x_{24}}{z_{13}z_{24}}-\frac{x_{14}x_{23}}{z_{14}z_{23}}
\quad .
\ea

The way to determine the coefficient 
\beq
C_5(j_1,j_2,j_3)=B(j_1)B(j_3)C \left ( \frac k2-j_1, j_2,
\frac k2-j_3 \right )\quad , \nonumber
\eeq
is reviewed in Appendix A where we also list some properties of $B(j)$ and
$C(j_1,j_2,j_3)$ which are useful for the calculations below. 

It can be verified that this result reduces to the four 
point function involving one spectral flow operator computed in reference 
\cite{malda3} when one of the generic spins vanishes. Indeed 
taking for instance $j_2=0$ 
in the five 
point function (\ref{5pt}), the coefficient $C_5$ gives 
$B(j_1)\delta(j_1-j_3)$ and the $x_i$, $z_i$ dependence  
reproduces the correlation function 
$\langle\Phi_{j_1}\Phi_{\frac k2}\Phi_{j_3}
\Phi_{j_4}\rangle$ (with the obvious change in labels) given in 
equation (5.25) of reference \cite{malda3}
when another  field has spin $\frac k2$. 
Actually taking $j_1=\frac k2$ in the equation computed by J. 
Maldacena and H. Ooguri, the $j_i$ dependent 
coefficient reduces to $B(j_3)\delta(j_3-j_4)$ and the coordinate dependence 
in both expressions matches, with the obvious renaming of spins and 
points.

As an intermediate step before computing the three point function we 
spectral flow once to obtain the following four point function
\beq
A_4^{w=1} = \langle
\Phi_{J_1,\bar J_1}^{w=1, j_1}
(x_1,z_1)
\Phi_{\frac k2}(x_2,z_2) 
\Phi_{j_2}(y_2,\zeta_2)\Phi_{j_3}(y_3,\zeta_3)\rangle \quad .
\eeq
This auxiliary result will also be useful for the computation of the 
four point function involving one spectral flowed and three unflowed
generic states which we perform in section 5.
Applying the prescription (\ref{1}) to $A_5$ and setting $y_1=x_1+t, 
\zeta_1=z_1+\epsilon$ we have to compute
\ba
A_4^{w=1} & = & \lim_{\epsilon, \bar\epsilon\rightarrow 0}
\epsilon^{m_1}\overline\epsilon^{\overline m_1}\int t^{j_1-m_1-1} \overline
 t^{j_1-\overline
m_1 -1} A_5(x_1,z_1,x_2,z_2,x_1+t, z_1+\epsilon,y_2,\zeta_2,y_3,\zeta_3)\; 
d^2 t\nonumber\\
& = & C_5(j_1,j_2,j_3)|x_{12}|^{2(j_1+j_2+j_3-k)}|\mu_1|^{2
(j_1-j_2-j_3)}
\lim_{\epsilon,\bar\epsilon\rightarrow 0}\epsilon^{m_1}
\overline\epsilon^{\overline m_1}\int d^2t \; t^{j_1-m_1-1}\overline
 t^{j_1-\overline m_1-1}
\nonumber\\ 
&&\qquad\qquad\times \; \left 
|\frac{x_{15}(x_{21}-t)}{z_{15}(z_{21}-\epsilon)}-
\frac {tx_{25}}{\epsilon z_{25}}\right |^{2(j_2-j_1-j_3)}
\left 
|\frac {t x_{24}}{\epsilon z_{24}}-\frac {(x_{21}-t)x_{14}}{(z_{21}-\epsilon)
z_{14}}\right |^{2(j_3-j_1-j_2)} \quad ,\nonumber
\ea
where we have omitted the 
overall dependence on the worldsheet coordinates $z_i$.
The integrand can be easily taken to the form $t^p\bar 
t^{\bar p}|at+b|^{2q}
|ct+d|^{2r}$ and we can then use 
an identity found in reference  \cite{geronimo}
which we recall in Appendix B, equation (\ref{gero}), to perform the 
integration. 
Taking next the limit $\epsilon,\overline \epsilon
\rightarrow 0$ the $\epsilon,\overline\epsilon$ dependence 
cancels 
and we finally  obtain
\ba
&&A_4^{w=1} = 2i\pi (-1)^{m_1+\bar m_1} B(j_1)B(j_3)
C\left (\frac k2-j_1, j_2,
\frac k2-j_3 \right )
\gamma(j_2-j_1-j_3+1)
\nonumber\\
&&\qquad\qquad \times ~
\frac{
\Gamma(j_1-J_1+\frac k2)
\Gamma(j_3-j_2+\bar J_1-
\frac k2)}{\Gamma(1-j_1+\bar J_1-\frac k2)
\Gamma(j_2-j_3-J_1+
\frac k2+1)
}\nonumber\\
&&\qquad\qquad\times \quad  \left [  
{_2 F_1}(
j_1+j_2-j_3,j_1-J_1+\frac k2, j_2-j_3-J_1+\frac k2+1
;u )\right.\nonumber\\
&&\qquad\qquad\qquad \qquad \times ~ {_2 F_1}(
j_1+j_2-j_3,j_1-\bar J_1+\frac k2, j_2-j_3-\bar J_1+\frac k2+1
;\bar u ) 
\nonumber\\
&& \qquad  + \lambda u^
{j_3+J_1-j_2-\frac k2}\bar u^
{j_3+\bar J_1-j_2-\frac k2}
{_2 F_1} (
j_1+j_3-j_2, j_1+J_1-\frac k2, j_3+ J_1-j_2-\frac k2+1;u)
\nonumber\\
&&\qquad\qquad\qquad\qquad\left.\times ~ {_2 F_1} (
j_1+j_3-j_2, j_1+\bar J_1-\frac k2, j_3+\bar J_1-j_2-\frac k2+1;\bar u)
\right ] \nonumber\\
&&\qquad\qquad\times ~
x_{12}^{j_2+j_3-J_1-\frac k2}\bar x_{12}^{j_2+j_3-\bar J_1-\frac k2}
|x_{14}|^{-4j_2}
x_{15}^{j_2-j_3-J_1+\frac k2} \bar x_{15}^{j_2-j_3-\bar J_1+\frac k2} 
\nonumber\\
&&\qquad\qquad\times ~ x_{25}^{J_1-\frac k2-j_2-j_3}
\bar x_{25}^{\bar J_1-\frac k2-j_2-j_3}
z_{14}^{\Delta_3-\Delta_1^{w}-\Delta_2-\Delta_{k/2}}
\bar z_{14}^{\Delta_3-\bar \Delta_1^{w}-\Delta_2-\Delta_{k/2}}
z_{15}^{\Delta_2-\Delta_1^{w}-\Delta_3+\Delta_{k/2}}\nonumber\\
&&\qquad\qquad\times ~ 
\bar z_{15}^{\Delta_2-\bar \Delta_1^{w}-\Delta_3+\Delta_{k/2}}
z_{45}^{\Delta_{k/2}+\Delta_1^{w}-\Delta_3-\Delta_{2}}
\bar z_{45}^{\Delta_{k/2}+\bar \Delta_1^{w}-\Delta_3-\Delta_{2}}
|z_{25}|^{k}\nonumber\\
&&\qquad \qquad \times \quad 
z^{J_1}\bar z ^{\bar J_1}|1-u|^{2(j_1-j_2-j_3)}|1-z|^{-2j_2}\quad .
\label{a4wk}
\ea
Here $x=\frac 
{x_{12}x_{45}}{x_{14}x_{25}}, 
z=\frac {z_{12}z_{45}}{z_{14}z_{25}}$, $u=\frac{1-x}{1-z}$ and

\ba
\lambda & = & \frac{\gamma(j_1+j_3-j_2)\Gamma(j_2-j_3-J_1+\frac k2+1)
\Gamma(j_1+\bar J_1-\frac k2)}{\gamma(j_1+j_2-j_3)
\Gamma(j_3+\bar J_1-j_2-\frac k2+1)\Gamma(j_3+\bar J_1-j_2-\frac k2)}
\nonumber\\
&&\qquad \times ~ \frac{\Gamma(\bar J_1-\frac k2-j_1+1)
\Gamma(j_2-j_3-J_1+\frac k2)}{\Gamma(j_1-J_1+\frac k2)
\Gamma(-j_1-J_1+\frac k2+1)}\quad .
\ea

Now, the three point function (\ref{tre2}) can be obtained either spectral
flowing once more from this four point function or
fusing two physical fields in the five point function (\ref{5pt}), 
say  $\Phi_{j_{1}}(y_{1},\zeta_{1})$ 
and $\Phi_{j_2}(y_2,\zeta_2)$, with the spectral flow operators
 through the prescription $(\ref{1})$. The first procedure is useful to
determine the coordinate dependence of the three point function whereas the
second one is more convenient to obtain the spin dependent coefficient.
Therefore we present both.

Let us first start from the four point function (\ref{a4wk}). It is 
convenient to rename $x_2, z_2 \rightarrow x_3, z_3$ and
set $x_5\equiv y_3=x_3+s$ and $z_5\equiv \zeta_3=z_3+\xi$. 
By definition we have 

\ba
A_3^{w=1,w=1}& = &\langle\Phi_{J_1,\bar J_1}^{w=1,j_1}(x_1,z_1)
\Phi_{j_2}(y_2,\zeta_2) \Phi_{J_3,\bar J_3}^{w=1,j_3}(x_3,z_3)
\rangle
\nonumber\\
&=&C_5(j_1,j_2,j_3)\gamma(j_2-j_1-j_3+1)
\frac{
\Gamma(j_1-J_1+\frac k2)
\Gamma(j_3-j_2+\bar J_1-\frac k2)}{\Gamma(1-j_1+\bar J_1-\frac k2)
\Gamma(j_2-j_3-J_1+\frac k2+1)}
\nonumber\\
&\times &
2i(-1)^{m+\overline m}\pi x_{13}^{j_2+j_3-J_1-\frac k2}
\bar x_{13}^{j_2+j_3-\bar J_1-\frac k2}|x_{14}|^{-4j_2}\nonumber\\
&\times &z_{14}^{\Delta_3-\Delta_1^w-\Delta_2
-\Delta_{\frac k2}}
\bar z_{14}^{\Delta_3-\bar \Delta_1^w-\Delta_2
-\Delta_{\frac k2}}\nonumber\\
&\times &\lim_{\xi,\bar\xi\rightarrow 0}\xi^
{m_2+\frac k2}\bar\xi^{\overline m_2+\frac k2}  
(z_{13}-\xi)^{\Delta_2
-\Delta_3-\Delta_1^{w}+\Delta_{\frac k2}}
 (\bar z_{13}-\bar\xi)^{\Delta_2
-\Delta_3-\bar \Delta_1^{w}+\Delta_{\frac k2}}
\nonumber\\
&\times & 
(z_{43}-\xi )^{-\Delta_2
-\Delta_3+\Delta_1^{w}+\Delta_{\frac k2}}
(\bar z_{43}-\bar\xi)^{-\Delta_2
-\Delta_3+\bar \Delta_1^{w}+\Delta_{\frac k2}}z^{J_1}\bar z^{\bar J_1}
|1-z|^{-2j_2}
\nonumber\\
&\times & \int d^2 s ~  s^{j_3-m_3-1}\bar s^{j_3-\overline m_3-1}
s^{J_1-\frac k2-j_2-j_3}\bar s ^{\bar 
J_1-\frac k2-j_2-j_3}\nonumber\\
&&\times (x_{13}-s)^{j_2-j_3-J_1+\frac k2}(\bar x_{13}-\bar 
s)^{j_2-j_3-\bar 
J_1+\frac k2}
|1-u|^{2(j_1-j_2-j_3)} \nonumber\\
&&\times \left [  
{_2 F_1}(
j_1+j_2-j_3,j_1-J_1+\frac k2, j_2-j_3-J_1+\frac k2+1
;u ){\overline{_2F_1}}(\bar u)\right.\nonumber\\
&&\quad  + ~ \lambda ~ u^
{j_3+J_1-j_2-\frac k2} ~ \bar u^
{j_3+\bar J_1-j_2-\frac k2}\nonumber\\
&&\left. \qquad \times {_2 F_1} (
j_1+j_3-j_2, j_1+J_1-\frac k2, j_3+ J_1-j_2-\frac k2+1;u)
{\overline{_2F_1}}(\bar u)\right ] \quad , 
\nonumber\\
\ea
where ${\overline{_2F_1}}$ denotes  
the hypergeometric function in the previous factor with the replacement
$ J_1 \rightarrow \bar J_1$ in the arguments. It is convenient to 
change variables as suggested by the following definition
\beq
u = \frac{(x_{13}-s)x_{23}}{x_{12}s}\frac{\xi z_{12}}{z_{23}(z_{13}-\xi)} =
e\frac{x_{13}-s}{s} \quad .
\eeq 
Namely, taking 
$s=\frac{ex_{13}}{u+e}$ it can be shown that the exponents of 
$\xi,\bar\xi$  cancel
and the integral can be rewritten as
\ba
A_3^{w=1,w=1} &=& 
C_5(j_1,j_2,j_3)
\gamma(j_2-j_1-j_3+1)
\frac{
\Gamma(j_1-J_1+\frac k2)
\Gamma(j_3-j_2+\bar J_1-\frac k2)}{\Gamma(1-j_1+\bar J_1-\frac k2)
\Gamma(j_2-j_3-J_1+\frac k2+1)}
\nonumber\\
&\times &
2i\pi (-1)^{m+\overline m}
 x_{12}^{J_3-J_1-j_2}\bar x_{12}^{\bar J_3-j_2-\bar J_1}
x_{13}^{j_2-J_1-J_3}
\bar x_{13}^{j_2-\bar J_1-\bar J_3}
x_{23}^{J_1-j_2-J_3}\bar x_{23}^{\bar J_1-j_2-\bar J_3}
\nonumber\\
&\times & z_{12}^{\Delta_3^w
-\Delta_1^w-\Delta_2}\bar z_{12}^{\bar\Delta_3^w
-\bar\Delta_1^w- \Delta_2}
z_{23}^{-\Delta_2
-\Delta_3^w+\Delta_1^{w}}\bar z_{23}^{-\Delta_2
-\bar \Delta_3^w+\bar \Delta_1^{w}}
z_{13}^{\Delta_2-\Delta_1^w-\Delta_3^w}
\bar z_{13}^{\Delta_2-\bar \Delta_1^w-\bar\Delta_3^w}\nonumber\\
&\times & 
\lim_{\xi,\bar\xi\rightarrow 0}\int d^2 u~  u^{j_2-j_3-J_1+\frac k2}
\bar u^{j_2-j_3-\bar J_1+\frac k2}
\nonumber\\
&&\qquad \times ~
(u+e)^{J_3+j_2-J_1}(\bar u+\bar e)^{\bar J_3+j_2-\bar J_1}
|1-u|^{2(j_1-j_2-j_3)} \nonumber\\
&&\qquad \times ~ \left [  
{_2 F_1}(
j_1+j_2-j_3,j_1-J_1+\frac k2, j_2-j_3-J_1+\frac k2+1
;u ){\overline{_2F_1}(\bar u)}\right.\nonumber\\
&&\qquad\qquad   + ~ \lambda ~ u^
{j_3+J_1-j_2-\frac k2} ~ \bar u^
{j_3+\bar J_1-j_2-\frac k2}\nonumber\\
&&\left. \qquad \quad \times ~ {_2 F_1} (
j_1+j_3-j_2, j_1+J_1-\frac k2, j_3+ J_1-j_2-\frac k2+1;u)
{\overline{_2F_1}}(\bar u)\right ] ,
\nonumber
\ea
so that we can safely take the limit $\xi,
\bar\xi\rightarrow 0$ inside the integral. 

Thus we have found the functional form of the three point function but we still
have to determine the $j_i, J_i$ dependence. One way to do it is to solve the
integral above. This can be done using results that have been 
found in reference \cite{geronimo}, but it is very tedious. Therefore
we will proceed along the alternative path, $i.e.$
spectral flowing twice directly from the five point function. 

Since
the coordinate dependence of the three point function has been determined
 we can use
conformal invariance to fix three of the insertion points of the fields
in $A_5$ whereas the other two get fixed from the spectral flow operation.
In this way we find an integral which can be explicitly computed following
\cite{fuku, satoh}.

Let us fix $x_1=z_1=0,
x_2=z_2=1, y_3=\zeta_3=\infty$
and set 
$y_1=x_1+t_1, \zeta_1=z_1+\epsilon_1, y_2=x_2+t_2$ and $\zeta_2=z_2
+\epsilon_2$ in (\ref{5pt}).
Then the three point function takes the following form
\ba
&&\left<\Phi^{w 
=1,j_{1}}_{J_{1},{\bar J}_{1}}(0,0)\Phi_{J_{2},{\bar J}_{2}}^{w=1, 
j_2}(1,1)
\Phi_{j_{3}}(\infty,\infty)\right> 
=  \lim_{\epsilon_1,\epsilon_2,
 \bar\epsilon_1,\bar\epsilon_2\rightarrow 0}
\epsilon_1^{m_1}\overline\epsilon_1^{\overline m_1}\epsilon_2^{m_2}
\overline\epsilon_2^{\overline m_2}
\nonumber\\
&& \times ~ \int d^2 t_1 d^2 t_2\; 
t_1 ^{2(j_1-m_1-1)} \bar t_1 ^{2(j_1-\overline m_1-1)}
t_2^{2(j_2-m_2-1)} \bar t_2^{2(j_2-\overline m_2-1)} A_5(
t_1,\bar t_1, \epsilon_1, \bar\epsilon_1; t_2,\bar t_2, \epsilon_2,
\bar\epsilon_2)\nonumber\\
&&=  \lim_{
\epsilon_1,\epsilon_2,
 \bar\epsilon_1,\bar\epsilon_2\rightarrow 0}
\epsilon_1^{m_1}\overline\epsilon_1^{\overline m_1}\epsilon_2^{m_2}
\overline\epsilon_2^{\overline m_2}\int d^2 t_1 d^2 t_2 \;
t_1 ^{2(j_1-m_1-1)} \overline t_1 ^{2(j_1-\overline m_1-1)}
t_2^{2(j_2-m_2-1)} \overline t_2^{2(j_2-\overline m_2-1)} \nonumber\\
&&
\times ~ \left |\frac{\epsilon_1-t_1}{\epsilon_1(1+\epsilon_1)}
\right |^{2(j_2-j_1-j_3)}\left |\frac{\epsilon_2-t_2}{\epsilon_2(1+\epsilon_2)}
\right |^{2(j_1-j_2-j_3)}\left |\frac{t_1t_2}{\epsilon_1\epsilon_2}
-\frac{(1+t_1)(1-t_2)}{(1+\epsilon_1)(1-\epsilon_2)}
\right |^{2(j_3-j_1-j_2)}\nonumber\\
&&\qquad \times ~
|\epsilon_1(1+\epsilon_1)|^{-2j_1}|\epsilon_2(1+\epsilon_2)|^{-2j_2}
|1+\epsilon_1-\epsilon_2|^{2(\Delta_3-\Delta_1-\Delta_2)}\; .
\ea
Performing the change of variables $t_1=
u\epsilon_1$, $t_2=v\epsilon_2$, the exponents of $\epsilon_1$ and $\epsilon_2$
cancel and the integral becomes
\ba
&&\int d^2u\; d^2 v \;
u^{j_1-m_1-1}\bar u^{j_1-\overline m_1-1}\bar 
v^{j_2-m_2-1}v^{j_2-\overline 
m_2-1}
\nonumber\\ && \qquad\qquad\qquad\qquad\qquad\quad 
\times\; |u-1|^{2(j_2-j_1-j_3)} 
|v-1|^{2(j_1-j_2-j_3)}|uv-1|^{2(j_3-j_1-j_2)}\; .\nonumber
\ea
Defining $v^\prime = v^{-1}$ this
 takes the form of the
integral  computed in reference \cite{fuku} 
(see also \cite{satoh}) which we review in Appendix B for completeness. 
Reinserting the coordinate dependence, the final result is 
\ba
&&\left<\Phi^{w 
=1,j_{1}}_{J_{1},{\bar J}_{1}}(x_{1},z_{1})\Phi_{J_{2},{\bar J}_{2}}^{w=1, 
j_2}(x_{2},z_{2})
\Phi_{j_{3}}(x_{3},z_{3})\right> = B(j_1)B(j_2)C\left (\frac k2-j_1, 
\frac k2-j_2, j_3\right )\nonumber\\
&&\quad \times ~
\frac{\Gamma(j_3-J_1+J_2)\Gamma(j_3+\bar J_1-\bar J_2)
\Gamma(2-j_1-j_2-j_3)^2}{\Gamma(1-j_3-
J_1+ J_2)\Gamma(1-j_3+\bar J_1-\bar J_2)}
~ W(j_1,j_2,j_3,J_1,J_2,\bar J_1,\bar J_2)\nonumber\\
&&\quad \times ~ 
x_{12}^{j_3-J_1-J_2}
\bar x_{12}^{j_3-\bar J_1-\bar J_2}
x_{13}^{(J_2-J_1-j_3)}
\bar x_{13}^{\bar J_2-\bar J_1-j_3}
x_{23}^{J_1-J_2-j_3}
\bar x_{23}^{\bar J_1-\bar J_2-j_3} 
\nonumber\\
&&\quad \times ~ 
z_{12}^{\Delta_3-\Delta_1^{w=1}-\Delta_2^{w=1}}
\bar z_{12}^{\Delta_3-\bar \Delta_1^{w=1}-\bar \Delta_2^{w=1}}
z_{13}^{\Delta_2^{w=1}-\Delta_1^{w=1}-\Delta_3}
\bar z_{13}^{\bar \Delta_2^{w=1}-\bar \Delta_1^{w=1}-\Delta_3}\nonumber\\
&&\quad \times ~
z_{23}^{\Delta_1^{w=1}
-\Delta_2^{w=1}-\Delta_3}
\bar z_{23}^{\bar \Delta_1^{w=1}-\bar \Delta_2^{w=1}-\Delta_3}\; ,
\label{3pww}
\ea
where
\ba
&&W(j_i,J_i,\bar J_i) ~ =~ s(j_2-j_1-j_3) 
G\left [
\begin{array}{c}
j_2 + J_2-\frac k2,j_2-j_1-j_3+1, 1-j_3+J_2-J_1\\
j_2-j_1+J_2-J_1+1, 2-j_1-j_3+J_2-\frac k2
\end{array}\right ] \nonumber\\
&&\quad \qquad\qquad \times ~ \left \{ s(j_1-j_2-j_3)G\left [
\begin{array}{c}
j_1-j_2 -j_3+1, j_1+\bar J_1-\frac k2, 1-j_3+\bar J_1-\bar J_2\\
2-j_2-j_3+\bar J_1-\frac k2, j_1-j_2+\bar J_1-\bar J_2+1
\end{array}\right ]\right.\nonumber\\
&&\quad \qquad\qquad \qquad \left. - ~
s(1-2 j_2) 
G\left [\begin{array}{c}
j_2 -\bar J_2+\frac k2, j_2-j_1-j_3+1, 1-j_3+\bar J_1-\bar J_2\\
j_2-j_1+\bar J_1-\bar J_2+1, 2-j_1-j_3-\bar J_2+\frac k2
\end{array}\right ] \right \} \nonumber\\
&&\quad\qquad\qquad + ~ 
s(j_1-j_2-j_3) 
G\left [\begin{array}{c}
j_1-j_2 -j_3+1, j_1-J_1+\frac k2, 1-j_3- J_1+J_2\\
2-j_2-j_3- J_1+\frac k2, 1+j_1-j_2+ J_2-J_1
\end{array}\right ] \nonumber\\
&& \quad\qquad\qquad \times ~ \left \{ - s(1-2j_1)G\left [\begin{array}{c}
j_1-j_2 -j_3+1, j_1+\bar J_1-\frac k2, 1-j_3+\bar J_1-\bar J_2\\
2-j_2-j_3+ \bar J_1-\frac k2, j_1-j_2+ \bar J_1-\bar J_2+1
\end{array}\right ] \right.\nonumber\\
&&\quad\qquad\qquad \left.   + ~ s(j_2-j_1-j_3)
G\left [\begin{array}{c}
j_2 -\bar J_2+\frac k2, j_2-j_1-j_3+1, 1-j_3+\bar J_1-\bar J_2\\
j_2-j_1+\bar J_1-\bar J_2+1, 2-j_1-j_3-\bar J_2+\frac k2
\end{array}\right ] 
\right \}\; ,\nonumber\\
\ea
with 
$G\left [ \begin{array}{c}
a,b,c\\
e,f \end{array}
\right ]
\equiv \frac {\Gamma(a)\Gamma(b)\Gamma(c)}{\Gamma(e)\Gamma(f)}
{_3F_2}(a,b,c;e,f;1)$ and $s(a)=sin(\pi a)$. The expression for 
$W(j_i,J_i,\bar J_i)$  can be rewritten using standard
properties of the generalized
 hypergeometric function listed in Appendix B to show 
that, since $m_i-\overline m_i\in$ {\bf Z}, it is symmetric with respect
to $J_i$ and $\bar J_i$.

As a consistency check on the result (\ref{3pww}), we 
 verify that it reduces to 
the two point function of one unit spectral
flowed states when $j_3=0$. Indeed we obtain
\ba
&&\left<\Phi^{w 
=1,j_{1}}_{J,{\bar J}}(x_{1},z_{1})\Phi_{J,{\bar J}}^{w=1, 
j_2}(x_{2},z_{2})
\Phi_{j_{3}=0}(x_{3},z_{3})\right> = 
x_{12}^{-2J}\bar x_{12}^{-2\bar J}
z_{12}^{-2\Delta_1^{w=1}}\bar z_{12}^{-2\bar\Delta_1^{w=1}}
\nonumber\\
&&\qquad \quad \times B(j_1) \delta(j_1-j_2)
\Gamma(0)\frac{\Gamma(1-2j_1)}{\Gamma(2j_1)}
\frac{\Gamma(j_1+m_1)\Gamma(j_1-\bar m_1)}
{\Gamma(1-j_1+m_1)\Gamma(1-j_1-\bar m_1)}\quad
,\nonumber
\ea
in agreement with the results in \cite{malda3}. Here we have used the 
properties of $B(j)$ and $C(j_1,j_2,j_3)$ which are listed
in Appendix A. 
We can identify the factor $\Gamma(0)$ in this expression with the 
volume of the conformal group of $S^2$ with two fixed points, namely
$V_{conf}=\int d^2 z|z|^{-2}$. Actually, as discussed in reference 
\cite{malda3}, in general a divergence arises when computing correlators 
which
include spectral flowed fields. The definition (\ref{1}) of these fields
already contains a rescaling $\tilde\Phi\rightarrow \Phi=V_{conf}\tilde\Phi$,
thus in this case there is a factor $V_{conf}^2$ whose product with $V_{conf}
^{-1}$ in the expression (5.13) of reference \cite{malda3}, computed in the
$m$ basis, explains the  factor $V_{conf}\sim \Gamma(0)$ which we have
found here.

Let us analyze the properties of our result. The function $W(j_i, 
J_i,\bar J_i)$ is analytic
in its arguments for states belonging to the continuous 
representation or their spectral flow images. Therefore the three point 
function (\ref{3pww}) is perfectly 
well behaved 
and finite for normalizable operators with $j=\frac 12 + is$, as expected.
If one of the original unflowed states, say $\Phi_{j_1}$, belongs  to a 
lowest 
weight representation, $i.e., m_1=j_1+n_1, \overline m_1 = j_1 + 
\overline n_1$ with $n_1, \bar n_1 =0,1,2\cdots$, then it can be 
shown that $W(j_i,J_i,\bar J_i)$ 
greatly simplifies, and taking further $n_1, 
\bar n_1 = 0$ the hypergeometric functions become unity. The analysis 
of $W(j_i,J_i,\bar J_i)$ completely agrees with that of reference \cite{satoh} 
(taking into account the change in notation). However notice that we are 
dealing with a winding conserving three point function which includes two 
one unit spectral flowed states whereas \cite{satoh} considers unflowed 
states. Moreover (\ref{3pww}) is an $x$ basis correlator 
unlike the $m$ basis expression analyzed in \cite{satoh}.

The three point function (\ref{3pww}) has various poles which come from 
the poles in $C_5$, in the $\Gamma-$functions and in the 
unrenormalized hypergeometric functions. $C_5$ has the same poles
as the unflowed three point function, namely at
\beq
j=n+m(k-2), \qquad -(n+1)-(m+1)(k-2), \qquad n,m=0,1,2,\cdots\quad ,
\eeq
with
\beq
j=1-j_1-j_2-j_3, \quad
j_1-j_2-j_3, \quad j_2-j_3-j_1, \quad j_3-j_2-j_1\; .
\eeq
The $\Gamma-$functions add the following poles  
\beq
J_1=J_2+j_3+n, \quad J_2=J_1+j_3+n \quad ,
\label{polosJ}
\eeq
and similar ones for $\bar J_1, \bar J_2$.
The poles of $G\left [ \begin{array}{c}
a,b,c\\
e,f \end{array}\right ]$ are at $a, b, c, u = -n$, with $u=e+f-a-b-c$,
and thus they are all contained in the previous ones except for
the poles signaling the 
presence of spectral flowed images of the discrete representations, $e.g.
~ m_1=j_1+n_1, \overline m_1=j_1+\overline n_1$.
Therefore the pole structure is as discussed in reference \cite{malda3} in 
the unflowed case with the addition of (\ref{polosJ}), which are analogous to
poles in the $S$ matrix of string theory in Minkowski space.

\section{Ward identities, KZ and null vector equations}

The computation of more complicated correlation functions
along the lines of the previous section would require to start from higher 
point amplitudes.
Actually the cases following in complexity, namely  
the three point
 function including three one-unit spectral flowed operators or
the four point function involving one $w=1$ field
require the knowledge of the six 
point function with three spectral flow operators and three physical states 
or the five point function with one $\Phi_{\frac k2}$ and four generic
unflowed fields respectively.
In this section we discuss general properties of correlation functions
containing $w=1$ spectral flowed operators in the $x$ basis
in order to find an alternative method to compute such more complicated 
amplitudes. More
specifically, we will derive the Ward identities and the modified KZ and
null vector equations to be satisfied by generic correlators including 
$w=1$ fields. We begin by 
giving an account of the already known results on the subject and the 
difficulties we expect to find in order to make further progress.

\subsection{Ward identities}

We investigate first the form of the Ward identities when the
definitions (\ref{1}) or (\ref{1a}) are used for the $w=1$ field. 
This can also be understood as an additional consistency check on such  
definition.

Let us start by considering $N$ point functions of 
primary $w=0$ fields,
\ba
A_{N}\equiv\langle 
\Phi_{j_{1}}(x_{1},z_{1})\Phi_{j_2}(x_{2},z_{2})
\cdots \Phi_{j_{N}}(x_{N},z_{N})\rangle\; .\nonumber
\ea
It is well known that the global SL(2) symmetry of the WZW model 
determines the Ward identities to be satisfied by the correlation 
functions, namely
\ba
0 &=& \sum_{i=1}^{N}\frac{\partial A_{N}}{\partial x_{i}}\; 
,\label{wix1}\\
0 &=& \sum_{i=1}^{N}\left(x_{i}\frac{\partial}{\partial 
x_{i}}+j_{i}\right)\;A_{N}\; , \label{wix2}\\
0 &=& \sum_{i=1}^{N}\left(x_{i}^{2}\frac{\partial}{\partial
x_{i}}+2j_{i}x_{i}\right)\;A_{N}\; . 
\label{2}
\ea

Now suppose we consider an $N+1$ point function including one spectral 
flow 
operator $\Phi_{\frac k2}$ at position $(x_2, z_2)$, 
\ba
A_{N+1}\equiv\langle 
\Phi_{j_{1}}(x_{1},z_{1})\Phi_{\frac k2}(x_{2},z_{2})
\cdots \Phi_{j_{N+1}}(x_{N+1},z_{N+1})\rangle\; ,
\label{an+1}
\ea
and take
\beq
x_{1}=x_{2}+y\; ,\qquad\qquad z_{1}=z_{2}+\epsilon\; .
\label{4}
\eeq
In order to obtain the differential equations determined by the Ward 
identities
one has to be careful that the derivatives  act 
only once on each of the points where the fields are inserted. Therefore 
it is convenient to perform the following transformation
\beq
\frac{\partial}{\partial x_{1}}\longrightarrow \frac{\partial}{\partial 
y}\; 
,\qquad\qquad \frac{\partial}{\partial x_{2}}\longrightarrow 
\frac{\partial}{\partial x_{2}}-\frac{\partial}{\partial y}\; , 
\label{5}
\eeq
so that the derivatives with respect to $x_2$ act only on the field in the 
second position and not on the first one.
The Ward identities transform accordingly, so for instance equation 
(\ref{wix2}) 
reads
\ba
0 &=& \sum_{i=1}^{N+1}\left(x_{i}\frac{\partial}{\partial
x_{i}}+j_{i}\right)\;A_{N+1}\nonumber\\ &=& 
\left[(x_{2}+y)\frac{\partial}{\partial 
y}+x_{2}\left(\frac{\partial}{\partial 
x_{2}}-\frac{\partial}{\partial 
y}\right)+j_{1}+\frac{k}{2}+\sum_{i=3}^{N+1}\left(x_{i}\frac{\partial}{\partial
x_{i}}+j_{i}\right)\right]\;A_{N+1}\nonumber\\
&=& \left[y\frac{\partial}{\partial
y}+x_{2}\frac{\partial}{\partial
x_{2}}+j_{1}+\frac{k}{2}+\sum_{i=3}^{N+1}\left(x_{i}\frac{\partial}{\partial
x_{i}}+j_{i}\right)\right]\;A_{N+1}
\; .
\label{6}
\ea
We want to derive the Ward identities for amplitudes containing $w=1$  
spectral flowed operators. Equation  (\ref{1}) suggests
to apply the following operation on (\ref{6})
\beq
\lim_{\epsilon, \bar\epsilon \rightarrow 0} \epsilon^{m}
{\bar\epsilon}^{{\bar m}}
\int d^2 y\; 
y^{j_{1}-m-1}{\bar y}^{j_{1}-{\bar m}-1} \; .
\label{oper}
\eeq
Integrating by parts, it can be seen that the 
Ward identity (\ref{6}) turns into 
\ba
0 &=& \left[-(j_{1}-m)+x_{2}\frac{\partial}{\partial
x_{2}}+j_{1}+\frac{k}{2}+\sum_{i=3}^{N+1}\left(x_{i}\frac{\partial}{\partial
x_{i}}+j_{i}\right)\right]\;A^{w}_{N}\nonumber\\
&=&\left[x_{2}\frac{\partial}{\partial
x_{2}}+\left(m+\frac{k}{2}\right)
+\sum_{i=3}^{N+1}\left(x_{i}\frac{\partial}{\partial
x_{i}}+j_{i}\right)\right]\;A^{w}_{N}\; ,
\label{7}
\ea
where from the definition (\ref{1}) we identify
\ba
A_{N}^w\equiv\langle 
\Phi_{m+\frac{k}{2},\;{{\bar m}+\frac{k}{2}}
}^{w=1,j_1}(x_{2},z_{2})\Phi_{j_3}(x_{3},z_{3})
\cdots \Phi_{j_{N+1}}(x_{N+1},z_{N+1})\rangle\; .
\label{ui}
\ea

Notice that equation (\ref{7}) is precisely of 
the same form as (\ref{wix2}) 
with the identification (\ref{1aa}) for the spin of the spectral flowed 
field. It can be shown that the same 
procedure gives an equivalent result for the two other Ward 
identities (\ref{wix1}) and (\ref{2}).

Now we focus on global conformal invariance, which determines the 
following differential equations for the correlators
\ba
0 &=& \sum_{i=1}^{N}\frac{\partial A_{N}}{\partial z_{i}}\; 
,\label{wiz1}\\
0 &=& \sum_{i=1}^{N}\left(z_{i}\frac{\partial}{\partial
z_{i}}+\Delta_{i}\right)\;A_{N}\; ,\label{wiz2}\\
0 &=& \sum_{i=1}^{N}\left(z_{i}^{2}\frac{\partial}{\partial
z_{i}}+2\Delta_{i}z_{i}\right)\;A_{N}\; ,
\label{9}
\ea
where the factors $\Delta_{i}$ are the conformal dimensions of the fields 
(see (\ref{wo})). 
 In order to derive the corresponding Ward identities for correlation 
functions containing one $w=1$ state we repeat the steps 
discussed above with the change of variables (\ref{4}) and
the corresponding transformation for the derivatives

\beq
\frac{\partial}{\partial z_{1}}\longrightarrow \frac{\partial}{\partial
\epsilon}\;
,\qquad\qquad \frac{\partial}{\partial z_{2}}\longrightarrow
\frac{\partial}{\partial z_{2}}-\frac{\partial}{\partial \epsilon}\; .
\label{11}
\eeq
Then for instance Eq.(\ref{wiz2}) becomes
\ba
0 &=& \sum_{i=1}^{N+1}\left(z_{i}\frac{\partial}{\partial
z_{i}}+\Delta_{i}\right)\;A_{N+1}\nonumber\\ &=&
\left[(z_{2}+\epsilon)\frac{\partial}{\partial
\epsilon}+z_{2}\left(\frac{\partial}{\partial
z_{2}}-\frac{\partial}{\partial
\epsilon}\right)+\Delta_{1}-\frac{k}{4}
+\sum_{i=3}^{N+1}\left(z_{i}\frac{\partial}{\partial
z_{i}}+\Delta_{i}\right)\right]\;A_{N+1}\nonumber\\
&=& \left[\epsilon\frac{\partial}{\partial
\epsilon}+z_{2}\frac{\partial}{\partial
z_{2}}+\Delta_{1}-\frac{k}{4}+
\sum_{i=3}^{N+1}\left(z_{i}\frac{\partial}{\partial
z_{i}}+\Delta_{i}\right)\right]\;A_{N+1}
\; .
\label{12}
\ea
Applying to this equation the operation
\beq
\lim_{y,\bar y\rightarrow 0}
y^{j_{1}-m}{\bar y}^{j_{1}-{\bar m}} 
\int d^{2}\epsilon\;
\epsilon^{m-1}{\bar\epsilon}^{{\bar m}-1}\; , 
\eeq
which is suggested by Eq.(\ref{1a}), and performing an integration by 
parts,
we can see that $A_N^w$ satisfies the spectral flowed Ward identity

\ba
0 = \left[z_{2}\frac{\partial}{\partial
z_{2}}+\left(\Delta_{1}-m-\frac{k}{4}\right)
+\sum_{i=3}^{N+1}\left(z_{i}\frac{\partial}{\partial
z_{i}}+\Delta_{i}\right)\right]\;A^{w}_{N}\; ,
\nonumber
\ea
which is of the same form as (\ref{wiz2}) with the following 
identification for the conformal dimension of the $w=1$ field 
\beq
\Delta_{1}^{w=1}=\Delta_{1}-m-\frac{k}{4}=\Delta_{1}-J+\frac{k}{4}\; ,
\label{13a}
\eeq
in agreement with (\ref{520}). A similar expression can be found for 
${\bar\Delta}_{1}^{w=1}$ in terms of ${\bar J}$. Again, all this goes 
through for the two other equations (\ref{wiz1}) and 
(\ref{9}). 

Therefore we conclude that the Ward identities to be satisfied by 
correlation functions 
including the operator $\Phi_{J, \bar J}^{w=1, j}$ coincide with those of 
the unflowed case with the modifications (\ref{1aa}) and (\ref{13a}) for 
the 
spin and conformal weight of the $w=1$ field respectively. This analysis 
can be generalized to correlation functions including an arbitrary number 
of $w=1$ states. From here the general form of the two and three point 
functions containing $w=1$ fields is completely determined, whereas 
the four point functions depend, as usual, on the anharmonic ratios. 

\subsection{Modified KZ and null vector equations}

Now we want to determine the form that the KZ and null vector equations 
take for correlators including $w=1$ fields. In order to do this, let us 
consider again the $N+1$-point function (\ref{an+1}). Consider $e.g.$ any 
point $z_{i}$ with $i\geq 3$. The correlator $A_{N+1}$ satisfies the 
standard 
KZ equation of the form

\ba
(k-2)\;\frac{\partial A_{N+1}}{\partial z_{i}}
&=&\sum_{n=1,\; n\not= 
i}^{N+1}\frac{1}{z_{i}-z_{n}}{\Bigg [}(x_{n}-x_{i})^{2}
\frac{\partial^{2}}{\partial x_{i}\partial x_{n}}+\nonumber\\ 
&&\qquad\qquad\quad +\;
2(x_{n}-x_{i})\left(j_{n}\frac{\partial}{\partial 
x_{i}}-j_{i}\frac{\partial}{\partial
x_{n}}\right)-2j_{i}j_{n}
{\Bigg ]}\; A_{N+1}\; .\nonumber\\
\label{15b}
\ea
In addition, since the spectral flow operator at $(x_{2},z_{2})$ 
has a null descendant, namely
$J^{-}_{-1}|j=k/2;m=k/2\rangle=0$,
then $A_{N+1}$ must also obey the following null vector equation
\beq
0=\sum_{n=1,\; n\not=
2}^{N+1}\frac{x_{n}-x_{2}}{z_{2}-z_{n}}\left[(x_{n}-x_{2})
\frac{\partial}{\partial x_{n}}+2j_{n}\right]\; A_{N+1}\; .
\label{16b}
\eeq
Our aim here is to perform similar manipulations to 
those in the
previous subsection, 
in order to investigate the form of the
equations to be satisfied by the $N$ point function including one $w=1$ 
field, namely $A_{N}^w$ in (\ref{ui}). The
general idea is that (\ref{ui}) can be obtained from (\ref{an+1}) by
performing the fusion of $\Phi_{j_{1}}(x_{1},z_{1})$ with
$\Phi_{k/2}(x_{2},z_{2})$ through the prescription $(\ref{1})$. In that
way, the equations to be satisfied by $A_{N+1}$, namely (\ref{15b}) and
(\ref{16b}), are expected to turn into those
to be obeyed by $A^{w}_{N}$. 

In order to do this, let us start by performing 
the 
change of variables (\ref{4})
in the KZ equation (\ref{15b}) which can then be rewritten as
\ba
(k-2)\;\frac{\partial A_{N+1}}{\partial z_{i}}&=& 
\frac{1}{z_{i}-z_{2}-\epsilon}{\Bigg [} 
(x_{2}+y-x_{i})^{2}\frac{\partial^{2}}{\partial x_{i}\partial 
y}\nonumber\\
&&\qquad +\; 2\;(x_{2}+y-x_{i})\left(
j_{1}\frac{\partial}{\partial x_{i}}-j_{i}\frac{\partial}{\partial y}
\right)
-2j_{i}j_{1}
{\Bigg ]}\; A_{N+1}\nonumber\\
&+&
\frac{1}{z_{i}-z_{2}}{\Bigg [}
(x_{2}-x_{i})^{2}\left(\frac{\partial^{2}}{\partial x_{i}\partial
x_{2}}-
\frac{\partial^{2}}{\partial x_{i}\partial
y
}\right)\nonumber\\
&&\qquad +\; 2\;(x_{2}-x_{i})\left(
\frac{k}{2}\;\frac{\partial}{\partial x_{i}}-j_{i}\frac{\partial}{\partial 
x_{2}}+j_{i}\frac{\partial}{\partial y}
\right)
-kj_{i}
{\Bigg ]}\; A_{N+1}\nonumber\\
&+&
\sum_{n=3,\; n\not=
i}^{N+1}\frac{1}{z_{i}-z_{n}}{\Bigg [}(x_{n}-x_{i})^{2}
\frac{\partial^{2}}{\partial x_{i}\partial x_{n}}+\nonumber\\
&&\qquad\qquad\quad +\;
2(x_{n}-x_{i})\left(j_{n}\frac{\partial}{\partial
x_{i}}-j_{i}\frac{\partial}{\partial
x_{n}}\right)-2j_{i}j_{n}
{\Bigg ]}\; A_{N+1}\; .\nonumber\\
\label{19b}
\ea

Now acting with the operator (\ref{oper}) and integrating by parts in $y$, 
we 
obtain the modified KZ equation for $A_N^w$, namely

\ba
(k-2)\;\frac{\partial A^{w}_{N}(J)}{\partial z_{i}}&=&
-\left(j_{1}-J+\frac{k}{2}-1\right)\;\frac{x_{2}-x_{i}}{(z_{i}-z_{2})^{2}}\;
\left[(x_{2}-x_{i})\frac{\partial}{\partial x_{i}}-2j_{i}\right]\; 
A^{w}_{N}(J+1)\nonumber\\
&+&
\frac{1}{z_{i}-z_{2}}{\Bigg [}(x_{2}-x_{i})^{2}
\frac{\partial^{2}}{\partial x_{i}\partial x_{2}}+\nonumber\\
&&\qquad\qquad\quad +\;
2(x_{2}-x_{i})\left(J\frac{\partial}{\partial
x_{i}}-j_{i}\frac{\partial}{\partial
x_{2}}\right)-2j_{i}J
{\Bigg ]}\; A^{w}_{N}(J)\nonumber\\
&+&
\sum_{n=3,\; n\not=
i}^{N+1}\frac{1}{z_{i}-z_{n}}{\Bigg [}(x_{n}-x_{i})^{2}
\frac{\partial^{2}}{\partial x_{i}\partial x_{n}}+\nonumber\\
&&\qquad\qquad\quad +\;
2(x_{n}-x_{i})\left(j_{n}\frac{\partial}{\partial
x_{i}}-j_{i}\frac{\partial}{\partial
x_{n}}\right)-2j_{i}j_{n}
{\Bigg ]}\; A^{w}_{N}(J)\; ,\nonumber\\
\; .
\label{al2}
\ea
where $J$ is the spin of the $w=1$ field, given by (\ref{1aa}).

The notation $A^{w}_{N}(J+1)$ indicates that we must replace 
$J\longrightarrow J+1$ in $A^{w}_{N}$. Thus, Eq.(\ref{al2}), which is 
interpreted as the KZ equation for an $N$ point function including one 
$w=1$ 
field, differs from the standard KZ equation for correlators of unflowed 
fields. In fact, notice that (\ref{al2}) is an {\it iterative} 
relation in the 
spin of the spectral  flowed field. As we will see, the property of being 
iterative 
in the spins of the spectral flowed fields will be common to all the 
equations to be satisfied by  correlators including fields in $w\not=0$ 
sectors. In fact, such a novel feature is not surprising, since it is 
inherited from the primary state condition (\ref{current}). We also point 
out that an equation analogous to (\ref{al2}) holds for the 
antiholomorphic part, where the 
iterative variable is ${\bar J}$ (see (\ref{1aa})).

Now, following a similar procedure with the null vector equation 
(\ref{16b}) we obtain an additional iterative equation, namely
\ba
\left(j_{1}+J-\frac{k}{2}-1\right)\;A^{w}_{N}(J-1)=
\sum_{n=3}^{N+1}\frac{x_{n}-x_{2}}{z_{2}-z_{n}}\left[
(x_{n}-x_{2})\frac{\partial}{\partial x_{n}}+2j_{n}
\right]\;
A^{w}_{N}(J)\; ,
\label{al4}
\ea
which is understood as the modified null vector equation to be 
satisfied by correlators containing one $w=1$ field. It supplements 
(\ref{al2}), so that both equations must be solved in order to find the 
explicit expression of $A^{w}_{N}$. As before, an analogous equation 
holds for the antiholomorphic part, with ${\bar J}$ as the 
iterative variable. The procedure detailed here can be 
extended to the case of correlators including any number of $w=1$ fields, 
where the spins of all the spectral flowed fields turn out to be iterative 
variables.
In the following section we will consider some specific calculations.

\section{Four point function including one $w=1$ field}

The purpose of this section is to explicitly solve the modified KZ and 
null vector equations corresponding to the four point function 
involving one $w=1$ field, namely
\beq
A_{4}^{w}=\left<\Phi_{j_{1}}(x_{1},z_{1})\Phi_{j_{2}}(x_{2},z_{2})
\Phi^{w=1,j_{3}}_{J,{\bar 
J}}(x_{3},z_{3})\Phi_{j_{4}}(x_{4},z_{4})\right> \; ,
\label{l1}
\eeq
which, according to the prior discussions, can be obtained from the 
five point function
\beq
\left<\Phi_{j_{1}}(x_{1},z_{1})\Phi_{j_{2}}(x_{2},z_{2})
\Phi_{j_{3}}(y,\zeta)
\Phi_{\frac{k}{2}}(x_{3},z_{3})\Phi_{j_{4}}(x_{4},z_{4})\right> 
\; ,
\label{l02}
\eeq
through the prescription (\ref{1}).

From the results of the previous section, we expect that $A_{4}^{w}$ had 
the same functional form as an unflowed four point function, but with the 
spin and conformal dimension of the $w=1$ field given by
$J=m +\;\frac{k}{2},\;
\Delta^{w=1}_{3}=\Delta_{3}-J+\frac{k}{4}$.
Thus, we consider the following expression for $A_{4}^{w}$
\ba
A_{4}^{w}&=&\int dj\; B(j_{3})
C(j_{1},j_{2},j)\; B(j)^{-1}
C\left(j,\frac{k}{2}-j_{3},j_{4}\right)
\nonumber\\ &\times&
D_{1}(j_{1},j_{2},j_{3},J,j_{4},j)\;
D_{2}(j_{1},j_{2},j_{3},{\bar
J},j_{4},j)\;
{\cal F}(z,x)\; {\bar{\cal F}}({\bar z},{\bar 
x})\;
\nonumber\\&\times&
\left(x_{43}^{j_{1}+j_{2}-j_{4}-J}
x_{42}^{-2j_{2}}x_{41}^{J+j_{2}-j_{4}-j_{1}}
x_{31}^{j_{4}-j_{1}-j_{2}-J}\right)
\nonumber\\
&\times &
\left(z_{43}^{\Delta_{1}+\Delta_{2}-\Delta_{4}-\Delta^{w=1}_{3}}
z_{42}^{-2\Delta_{2}}z_{41}^{\Delta^{w=1}_{3}+\Delta_{2}
-\Delta_{4}-\Delta_{1}}
z_{31}^{\Delta_{4}-\Delta_{1}-\Delta_{2}-\Delta^{w=1}_{3}}\right)
\nonumber\\ &\times & ({\rm antiholomorphic\; part})
\; ,
\label{4pw}
\ea
where the dependence in the coefficients $B$ and $C$ is inherited from the 
five point function (\ref{l02}) (see details in Appendix A). Notice that, 
due to the presence of the spectral flow operator, we 
are left with only one state in one of the two intermediate channels. The 
other one contributes the integral in $j$.

In addition, $D_{1}$ and $D_{2}$ are the parts of the coefficient of the 
four point function 
depending respectively 
on the right and left spins of 
the string states, whereas ${\cal F}$ 
and ${\bar{\cal F}}$ are functions of the cross ratios
\beq
z=\frac{z_{21}z_{43}}{z_{31}z_{42}}\; ,\qquad 
x=\frac{x_{21}x_{43}}{x_{31}x_{42}}\; .
\label{l24}
\eeq

Now plugging (\ref{4pw}) into the modified KZ and null vector equations 
(which, up to the obvious change in labels, are given by (\ref{al2}) and
(\ref{al4})) we 
find respectively 
the following iterative expressions \footnote{This is the modified KZ 
equation at ($x_1, z_1$).}
\ba
&&\left(j_{3}-J+\frac{k}{2}-1\right)\;\left[(1-x)\frac{\partial}{\partial 
x}+(j_{1}+j_{4}-j_{2}-J-1)\right]\; z\; D_{1}(J+1)\; {\cal 
F}_{J+1}\nonumber\\
&=& {\Bigg \{ }-(k-2)z(1-z)\frac{\partial}{\partial z}\; +\; 
x(1-x)(z-x)\frac{\partial^{2}}{\partial x^{2}}\;-
\;{\Bigg [}
(j_{4}-j_{1}-j_{2}+J-1)z
\nonumber\\ &&
\qquad\qquad\quad -2(j_{4}-j_{2}-1)xz+2(j_{1}+j_{2})x+
(j_{4}-j_{1}-3j_{2}-J-1)x^{2}
{\Bigg ]}\;\frac{\partial}{\partial x}\nonumber\\
&& \qquad\qquad 
+\left[2j_{2}(j_{4}-j_{2})-j_{3}(j_{3}-1)+J(J-1)-(k-2)J+k(k-2)/4\right]\; 
z \nonumber\\ &&\qquad\qquad\qquad\qquad\qquad\qquad\qquad
-2j_{2}(j_{4}-j_{1}-j_{2}-J)x-2j_{1}j_{2}
{\Bigg \} }\; D_{1}(J)\; {\cal F}_{J}\; ,\nonumber\\
\label{d1}
\ea
and
\ba
&&\left(j_{3}+J-\frac{k}{2}-1\right)\;(1-z)\; D_{1}(J-1)\; {\cal
F}_{J-1}\nonumber\\ && \quad =\;
\left[(j_{2}-j_{4}-j_{1}+J)(1-z)-2j_{2}(1-x)+(1-x)(z-x)\;
\frac{\partial }{\partial x}\;\right]\; D_{1}(J)\; 
{\cal F}_{J}\; .\nonumber\\
\label{d1j}
\ea
Here  $D_{1}(J\pm 1)$ (${\cal F}_{J\pm 1}$) indicates that 
we must replace $J\longrightarrow J\pm 1$ in $D_{1}$ (${\cal F}$). 
 Analogous expressions can be found for the 
antiholomorphic part $D_{2}(\bar J)\; ({\bar {\cal F}}_{\bar J}$).

Now we follow a similar route to that in the unflowed case 
\cite{malda3, tesch2} and expand ${\cal F}$ in powers of $z$ as 
follows
\beq
{\cal
F}(z,x)=z^{\Delta_{j}-\Delta_{j_{1}}-\Delta_{j_{2}}}\; x^{j-j_{1}-j_{2}}
\sum_{n=0}^{\infty}f_{n}(x)z^{n}\; .
\label{l4}
\eeq

We then focus on the lowest 
order of this expansion. We consider first the KZ equation
(\ref{d1}).  
The crucial result is that, to the lowest order in $z$, the iterative term 
in the l.h.s. 
does not contribute, whereas the r.h.s. reduces to an expression of 
precisely 
the same form as that of the lowest order of the standard (unflowed) KZ 
equation, as 
computed in 
\cite{tesch2}, with the only difference that $j_{3}$ is replaced by $J$. 
Thus we have the following solution\footnote{Actually the solution is a 
linear combination of the functions 
${_{2}F_{1}}(j-j_{1}+j_{2},j+J-j_{4},2j;x)$ and 
$x^{1-2j}{_{2}F_{1}}(1-j-j_{1}+j_{2},1-j+J-j_{4},2-2j;x)$. However, 
analogously as 
in the unflowed case \cite{malda3}, we may use the fact that, when 
inserted in (\ref{l4}), the two solutions are related to each other 
through the 
symmetry $j\longrightarrow 1-j$ which allows to keep only the first 
solution provided that in (\ref{4pw}) we now integrate $j$ over the entire 
imaginary axis, i.e. $\frac{1}{2}+i{\bf R}$.}
\beq
f_{0}={_{2}F_{1}}(j-j_{1}+j_{2},j+J-j_{4},2j;x)\; ,
\label{l5}
\eeq
where ${_2F_1}$ is the standard hypergeometric function.

Now we turn to the modified null vector equation (\ref{d1j}). Keeping the 
lowest order in $z$ and using (\ref{l5}) we find that the coefficient 
$D_{1}$ must satisfy the following iterative relation
\beq
\left(j_{3}+J-\frac{k}{2}-1\right)\; D_{1}(J-1)\; =\; (J-j-j_{4})\; 
D_{1}(J)\; ,
\label{l6}
\eeq
with an analogous expression for $D_{2}$. This 
allows to cancel the coefficient $D_1$ in (\ref{d1}) and (\ref{d1j})  
and we can write equations for 
all higher order terms in 
(\ref{l4}) starting from the lowest order (\ref{l5}), $i.e.$ we are able 
to 
find iterative equations for
$f_{n}$ in terms of $f_{n-1}$ (for $n\geq 1$). This is done by 
plugging (\ref{l4}) into (\ref{d1}) and (\ref{d1j}) and using (\ref{l6}). 
For
instance the modified KZ equation (\ref{d1}) gives
\ba
&& (j_{4}-J+j-1){\Bigg \{ }x^{2}(1-x)\frac{d^{2}}{dx^{2}}\; +\; 
\left[(j_{1}-j_{2}+j_{4}-J-2j-1)x+2j \right]x\frac{d}{dx}
\nonumber\\
&+& [(j_{1}+j_{2}-j)(2j_{2}-j_{4}+J+j)-2j_{2}(j_{1}+j_{2}-j_{4}+J)]x\; +\; 
(k-2)n 
{\Bigg \} }\; f^{(J)}_{n}\nonumber\\
&=&\left(j_{3}+J-\frac{k}{2}\right)\left(j_{3}-J+\frac{k}{2}-1\right)
\nonumber\\ &&\qquad\quad \times\;
\left[ (1-x)\frac{d}{dx}\; -\; (j_{1}+j_{2}-j)\frac{1-x}{x}\; +\; 
j_{1}-j_{2}+j_{4}-J-1
\right]\; f^{(J+1)}_{n-1}\nonumber\\
&+& (j_{4}-J+j-1){\Bigg \{ }
x(1-x)\frac{d^{2}}{dx^{2}}
\nonumber\\ &&\qquad\qquad\qquad +\;
\left[2(j_{1}+j_{4}-j-1)x-
(j_{1}+j_{2}+j_{4}+J-2j-1) 
\right]\frac{d}{dx}\nonumber\\
&&\qquad\qquad\qquad +\; (j_{1}+j_{2}-j)(j_{4}+J-j)\frac{1}{x}
-2j_{1}j_{4} - j_{3}(j_{3}-1)
\nonumber\\
&&\qquad\qquad\qquad +\;
2j(j_{1}+j_{4}-j)+J(J-1)+(k-2)(n-1-J+k/4)
{\Bigg \} }\; f^{(J)}_{n-1}\; .\nonumber\\ 
\label{l15}
\ea
An interesting thing about this equation is that, even when it is also 
iterative in $J$, as expected, such iterative terms are all related to the 
$(n-1)$-order factor, $f_{n-1}$, whereas there is no iterative term for 
$f_{n}$. This allows to write an equation for $f^{(J)}_{n}$ in terms of 
the data $f^{(J)}_{n-1}$ and $f^{(J+1)}_{n-1}$.

A similar procedure can be followed with the modified null vector 
equation (\ref{d1j}) and we obtain
\ba
&& \left[x(1-x)\frac{d}{dx}\; +\; (j_{1}-j_{2}-j)x\; +\;
j_{4}+j-J\right]\; f^{(J)}_{n}\; +\; (J-j-j_{4})\; f^{(J-1)}_{n}
\nonumber\\ &=&
\left[(1-x)\frac{d}{dx}\; +\; j_{4}+2j_{1}-j-J\; +\;
(j-j_{1}-j_{2})\;\frac{1}{x}\right]\; f^{(J)}_{n-1}\; +\; (J-j-j_{4})\;
f^{(J-1)}_{n-1}\; .
\nonumber\\
\label{l40}
\ea

The calculations we have performed so far are similar in spirit to those 
considered in the unflowed case  \cite{tesch2} (see also 
\cite{malda3}). Notice, however, that we have not completely determined 
the coefficient $D_{1}$ yet. Even though the functional dependence on the 
coordinates is fixed,
all the information we have to fully determine the spin dependent 
coefficient
 is the iterative expression 
(\ref{l6}). This means that the 
modified KZ and null vector equations do not completely specify the spin
dependence of the four point function.
 This is not surprising since a similar situation 
is found in the unflowed case. Nevertheless, we are still able to find a 
proper expression for the coefficient by requiring 
the following two 
 conditions: $i$) that it satisfies (\ref{l6}) (and a similar 
expression for $D_{2}$), and $ii$) that $A_4^w$ in (\ref{4pw}) 
correctly reduces to (\ref{3w}),
the three point function involving one spectral 
flowed field,  when one of the unflowed operators is the 
identity.

It can be shown that a solution to $i$) and $ii$) is 
given by
\beq
D_{1}D_{2}\sim
\frac{1}{\gamma\left(j_{1}+j_{2}+j_{3}+j_{4}-\frac{k}{2}\right)}\;
\frac{\Gamma\left(j_{3}+J-\frac{k}{2}\right)}
{\Gamma(1+J-j_{4}-j)}\;\frac{\Gamma(j_{4}+j-{\bar
J})}
{\Gamma\left(1-j_{3}-{\bar J}+\frac{k}{2}\right)}\; ,
\label{d1d2}
\eeq
up to a $k$ dependent coefficient. Requirement $i$) can be verified  
using 
standard properties of $\Gamma$-functions, whereas  $ii$) is 
also 
satisfied since, using the expression above together with (\ref{l5}) 
and 
some of the identities in Appendix A, it 
can be shown that (\ref{4pw}) reduces to (\ref{3w}) for $j_{2}=0$.

We should point out however that the solution (\ref{d1d2}) 
is not unique. For instance, the coefficient in 
(\ref{a4wk}) verifies both requirements $i$) and $ii$) but it
does not match (\ref{d1d2}).\footnote{When checking these last statements 
it must be taken 
into account that (\ref{a4wk}) involves, apart from the obvious changes in 
labels and the presence of a spectral flow operator, a result 
which is expressed in cross ratios other than those in 
(\ref{l24}), so that appropriate transformations must be performed on 
(\ref{a4wk}) before comparing it to the expressions in this section.}  
Such residual uncertainties might be removed studying the
factorization properties of the four point function (\ref{4pw}), following 
a similar 
path to that of section 4 in reference
\cite{malda3} for the unflowed case. However  
here the pole structure of the four point function presents additional 
difficulties since there are poles in the integral in the complex $j$ 
plane crossing the integration contour even before performing the analytic 
continuation. Therefore we leave this analysis for future work.

\section{Discussion and conclusions}

The purpose of this work was to study correlation functions 
involving one 
unit spectral flowed string states in AdS$_3$.
We have computed the three point function including one unflowed and two 
$w=1$ states in the WZW model in SL(2,R), and performed the 
analysis of the corresponding pole structure. We have also 
considered the 
four point function with one $w=1$ and three generic $w=0$ states.

We performed various 
checks  on our results. In particular, the Ward identities prescribing
the general form of correlation functions
 containing spectral flowed fields were discussed
and we then verified that the three and four
point functions computed indeed have the  form dictated by
conformal and global SL(2) invariance. In addition, we also verified 
that the three point function including two $w=1$ operators  
reproduces the
corresponding two point function of $w=1$ spectral flowed fields when the 
third operator is the
identity.

Our results represent one step forward 
towards establishing the consistency of
string theory on AdS$_3$. 
This would require 
the analysis of the 
factorization properties
of the four point function (\ref{4pw}).
 Actually the 
structure of the factorization of the unflowed
four point function contains several differences with the 
flat 
case and it would be interesting to see how they generalize when winding 
is considered.
Indeed, it was argued in reference \cite{malda3} that 
the four point functions do not factorize 
as expected into sums of products of three point functions with 
physical intermediate states unless the quantum numbers of the external 
states verify $j_1+j_2<\frac {k+1}2, j_3+j_4<\frac{k+1}2$. The 
interpretation of these constraints presented in \cite{malda3} indicates that
correlation functions violating these bounds do not represent well defined 
computations in the dual CFT description of the theory on the boundary. 
This explanation is similar to the interpretation of the upper bound on 
the spin of the physical states ($i.e., j<\frac{k-1}2$) as the condition 
that only local operators be considered in the boundary CFT. However in 
the later case one has a clear understanding of the constraint from the 
representations of SL(2,R) which define the theory in the bulk. Similarly 
one would like to better understand this unusual feature of the correlation 
functions from the worldsheet viewpoint. Moreover the factorization 
structure of the four point function (\ref{4pw}) would also be 
important to unambiguously determine the spin dependent coefficient 
$D_1D_2$.
However, as discussed in section 5,
 the factorization of $A_4^w$ 
presents additional 
difficulties to those encountered in the unflowed case since there are 
poles in the integral in the complex $j$ plane crossing the integration 
contour even before performing the analytic continuation. Therefore we postpone
the analysis of the consistency of string theory when
spectral flowed correlators are considered for future work.

Correlation functions involving states
in higher winding sectors have not been considered so far. This is an 
important ingredient for the complete determination of string theory on 
AdS$_3$. However this would require, along the lines we have presented 
here, first of all a proper definition of such fields in the $x$
basis, which is not known yet.

We have obtained the modified KZ and null vector equations to be satisfied 
by 
correlation functions containing $w=1$ spectral flowed 
fields.\footnote{See 
\cite{ribault} for a different approach to the
construction of the KZ equations for correlation functions containing
spectral flowed fields.} We 
have shown that
these are iterative equations relating amplitudes 
generically involving spectral flowed
fields with spins $J$, $J+1$ and $J-1$. We managed to manipulate these
expressions and analyze the four point function containing one $w=1$ 
field. The modified KZ and null vector equations  also allowed us to
obtain certain three point functions involving three
$w=1$ fields for particular combinations of spins
(see Appendix C). A similar analysis for the three point 
function containing two $w=1$ states shows, when comparing with the procedure
followed in section 3, that the specific spin relations found correspond to 
simplified
integrals in the spectral flowing procedure of the original higher point 
unflowed function involving the operators $\Phi_{\frac k2}$, and therefore
they seem to have no physical relevance. A more general {\it ansatz} than 
the one
we proposed in (\ref{27}), possibly involving hypergeometric functions, 
would be required in order 
to obtain the full three point function.

Finally let us observe that, 
according to the well known relation between correlation functions in 
the SL(2,{\bf R}) WZW model and Liouville theory \cite{zamo1, andreev},
our results can also be used to obtain amplitudes in this later theory. 
\footnote{See \cite{ribault, giribet} for  
more recent work on this connection.}

\section{Acknowledgments}

We would like to express our gratitude to S. Iguri and L. 
Nicol\'as for discussions and especially to J. Maldacena for
important suggestions on a previous version of this work
and for reading this manuscript.
E.H. is also grateful to V. S. Dotsenko for correspondence, and P.M. would 
like to thank J. Helayel-Neto for encouragement.
This work was supported by CLAF, CONICET, Universidad de Buenos
Aires and ANPCyT.

\section{Appendix A. Correlators containing
$\Phi_{\frac k2}$}

In this appendix we obtain the $j_i$ dependent coefficients for the 
five and six point functions respectively containing two and three 
spectral flow operators $\Phi_{\frac 
k2}$ as well as three operators of generic spins $j_1, j_2, j_3$, in the 
limit
where each $\Phi_{\frac k2}$ fuses with a $\Phi_j$ to give
a spectral flowed operator. That is, starting from

\beq
\langle \Phi_{j_1} \Phi_{\frac k2}\Phi_{j_2} 
\Phi_{\frac k2}\Phi_{j_3}\rangle \; ,
\eeq
and

\beq
\langle \Phi_{j_1} \Phi_{\frac k2}\Phi_{j_2}
\Phi_{\frac k2}\Phi_{j_3}\Phi_{\frac k2}\rangle \; ,
\eeq
we would like to respectively obtain the $j_i$ dependent coefficient of the 
following three point functions

\beq
\langle
\Phi^{w=1,j_1}_{J_1,\bar J_1}
\Phi^{w=1,j_2}_{J_2,\bar J_2}\Phi_{j_3}\rangle \; ,
\eeq
and

\beq
\langle 
\Phi^{w=1,j_{1}}_{J_{1},{\bar J}_{1}}
\Phi^{w=1,j_{2}}_{J_{2},{\bar J}_{2}}\Phi^{w=1,j_{3}}_{J_{3},\bar 
J_{3}}\rangle \; .
\eeq
In addition, we will perform analogous calculations for the five point 
function including only one $\Phi_{\frac k2}$, namely

\beq
\langle \Phi_{j_{1}} \Phi_{j_{2}}\Phi_{j_{3}}
\Phi_{\frac {k}{2}}\Phi_{j_{4}}\rangle \; ,
\eeq
in order to obtain the $j_i$ dependent coefficient of the four point 
function

\beq
\langle \Phi_{j_1} 
\Phi_{j_2}\Phi^{w=1,j_3}_{J,\bar J}\Phi_{j_4}\rangle \; .
\eeq 

The following properties of the $B$ and $C$ coefficients
of the two and three point functions will be useful, namely 
\cite{malda3}

\ba
B\left(\frac{k}{2}-j\right)&\sim& \frac{1}{B(j)}\; ,\label{a}\\
C\left(j_{1},j_{2},\frac{k}{2}\right )&\sim&
\delta\left(j_{1}+j_{2}-\frac{k}{2}\right )\; ,\label{d}\\
C\left(\frac{k}{2}-j,\frac{k}{2}-j,1\right)&\sim& \frac{1}{B(j)}\; 
,\label{e}\\
C(j_{1},j_{2},0) &=& B(j_{1})\delta(j_{1}-j_{2})\; ,\label{f}\\
C\left(\frac k2-j_1, \frac k2-j_2,j_3\right)&\sim& B\left(\frac 
k2-j_1\right)
B\left(\frac 
k2 - j_2\right)
C(j_1,j_2,j_3)\; ,
\label{62a}
\ea
where $\sim$ indicates that the 
identity holds up to a $k$ dependent ($j$ independent) factor. 

We start from the following formal expression for the 
OPE (see \cite{tesch2} for a detailed
definition of the OPE in the SL(2,C)/SU(2) WZW model)
\beq
\Phi_{j_1}(x_1,z_1) \Phi_{j_2}(x_2,z_2)
\sim \int d {j_i}\; Q(j_1,j_2,j_i)\Phi_{j_i}(x_2,z_2) \; ,
\label{ope}
\eeq 
where from now on we drop the $x_i,\; z_i$ dependent factors. The 
coefficient 
$Q$ can be determined
multiplying both sides of Eq.(\ref{ope}) by $\Phi_{j_3}$, namely 
\beq
\Phi_{j_1}(x_1,z_1) 
\Phi_{j_2}(x_2,z_2)\Phi_{j_3}(x_3,z_3)
\sim \int d {j_i}\; Q(j_1,j_2,j_i)\Phi_{j_i}(x_2,z_2) 
\Phi_{j_3}(x_3,z_3)\; ,
\label{ope3}
\eeq
and taking the
expectation values as
\beq
\langle\Phi_{j_1}(x_1,z_1)\Phi_{j_2}(x_2,z_2)\Phi_{j_3}(x_3,z_3)\rangle \sim
\int d{j}\; Q(j_1,j_2,j)\langle\Phi_{j}(x_2,z_2)\Phi_{j_3}(x_3,z_3)
\rangle\; .
\eeq

The two point
function $\langle\Phi_{j}\Phi_{j_3}\rangle$ in the right hand side gives 
two possible contributions which are proportional to
$\delta(j-j_3)$ and $\delta(j+j_3-1)$ (see Eq.(\ref{op})).
As discussed in reference \cite{satoh} they both give the same result for
$Q$, namely
\beq
Q(j_1,j_2,j_3)= \frac {C(j_1,j_2,j_3)}{B(j_3)} \; .
\eeq

Let us now repeat this procedure for the four point function. Starting from
Eq.(\ref{ope3}) we multiply both sides by $\Phi_{j_4}(x_4,z_4)$ and take 
the expectation value. One thus obtains 
\ba
\langle\Phi_{j_1}\Phi_{j_2}
\Phi_{j_3}
\Phi_{j_4}
\rangle  &\sim & 
\int d{j}\; Q(j_1,j_2,j)
\langle\Phi_{j}(x_2,z_2)\Phi_{j_3}(x_3,z_3)\Phi_{j_4}(x_4,z_4)
\rangle \nonumber\\
&\sim &\int dj\; C(j_1,j_2,j)\frac 1{B(j)} C(j,j_3,j_4)\; .
\label{4pt}
\ea

Suppose one of the fields
is a spectral flow operator, for instance $j_2=\frac k2$. The
 properties
(\ref{a}) and (\ref{d}) allow to perform the integral over $j$  and obtain 
the coefficient
 $B(j_1)C(\frac k2-j_1,j_3, j_4)$ for the three point function  
$\langle\Phi_{J,\bar J}^{w=1, j_1}\Phi_{j_3}\Phi_{j_4}\rangle$ 
\cite{malda3}. 

Similarly, if one starts from a five point function, the OPE (\ref{ope}) can be
used twice, say for $\Phi_{j_1}\Phi_{j_2}$ and $\Phi_{j_3}\Phi_{j_4}$,
with the result
\beq
\langle\Phi_{j_1}\Phi_{j_2}
\Phi_{j_3}
\Phi_{j_4}\Phi_{j_5}
\rangle  \sim 
\int d{j}\int dj^{\prime}\; Q(j_1,j_2,j)Q(j_3,j_4,j^\prime )
\langle\Phi_j(x_2,z_2)\Phi_{j^\prime }(x_4,z_4)\Phi_{j_5}(x_5,z_5)\rangle \; .
\nonumber\\
\eeq
Again, if there are spectral flow
operators, the result simplifies. For instance
consider $j_2=j_4=\frac k2$. In this case
one can perform the double integral above
 and obtain 
\beq
\langle\Phi_{j_1}\Phi_{\frac k2}
\Phi_{j_3}
\Phi_{\frac k2}\Phi_{j_5}
\rangle  \sim 
B(j_1)B(j_3)C\left(\frac k2-j_1, \frac k2-
j_3,j_5\right) \; ,
\label{parpar}
\eeq 
for the coefficient of the three point function 
$\langle\Phi_{J_1}^{w=1, j_1}\Phi^{w=1,j_3}_{J_3}\Phi_{j_5}\rangle$. 

In the more complicated case where we have only one $\Phi_{\frac k2}$, say 
$j_4=\frac k2$, the double integral turns into a single one of the form

\beq
\int dj\; \frac{B(j_{3})}{B(j)}\; C(j_{1},j_{2},j)\;
C\left(j,\frac{k}{2}-j_{3},j_{5}\right)\; ,
\eeq
which corresponds to the four point function 
$\left<\Phi_{j_{1}}\Phi_{j_{2}}
\Phi^{w=1,j_{3}}_{J}\Phi_{j_{5}}\right>$.

Finally if one starts from the six point function containing three spectral
flow operators and wants to obtain the coefficient for 
$\langle\Phi_{J_1}^{w=1, 
j_1}\Phi^{w=1,j_3}_{J_3}\Phi_{J_6}^{w=1,j_6}\rangle$,
 the OPE can be used three times and  the properties (\ref{a}) and (\ref{d})
determine the following corresponding coefficient
\beq
\langle\Phi_{j_1}\Phi_{\frac k2}
\Phi_{j_3}
\Phi_{\frac k2}\Phi_{\frac k2}\Phi_{j_6}
\rangle  \sim 
B(j_1)B(j_3)B(j_6)C\left(\frac k2-j_1, \frac k2-j_3,\frac k2-j_6\right) \; 
.
\label{parim}
\eeq

\section {Appendix B. Useful formulae}

In this appendix we collect some useful formulae that have been used in the
main body of the article. 

The following integral was found in \cite{geronimo} and it was used in 
section 3 in order to compute the four point function
$A_4^{w=1}$ involving one $w=1$ state, one spectral flow operator and two
generic unflowed states
\ba
&& \int d^2 t\; t^{p} \bar t^{\bar p}
\left |at+b \right |^{2q} \left | ct+d \right |
^{2r} ~ = ~2i(-1)^{p+\bar p}\pi \frac {\Gamma(p+1)\Gamma(q+1)\Gamma(-\bar p-
q-1)}{\Gamma(-\bar p)\Gamma(-q)\Gamma(p+q+2)} |d|^{2r} \nonumber \\
&& \quad \times ~ \frac {
b^{p+q+1} \bar 
b^{\bar p + q + 1}}{a^{p+1}\bar a^{\bar p + 1}}
\left [ 
{_2 F_1}\left 
(-r, 1+p, 2+p+q; \frac {cb}{ad}\right ){_2 F_1}\left 
(-r, 1+\bar p, 2+\bar p+q; \frac {\bar c \bar b}{\bar a \bar d}\right )
 \right. \nonumber\\
&&\qquad \qquad \left. + ~  \lambda ~
\left (\frac {cb}{ad}
\right )^{-1-p-q}\left (\frac {\bar c\bar b}{\bar a \bar d}
\right )^{-1-\bar p- q}
{_2 F_1}\left (-q,-1-p-q-r,-p-q;
\frac{cb}{ad}\right ) \right.\nonumber\\
&& \qquad \qquad\qquad \qquad
\times ~ \left. {_2 F_1}\left (-q,-1-\bar p-q-r,-\bar p-q;
\frac{\bar c \bar b}{\bar a \bar d}\right ) 
\right ]\quad ,
\label{gero}
\ea
where
\beq
\lambda = \frac{\Gamma(p+q+2)\Gamma(-q-\bar p-r-1)
\gamma(-q)\Gamma(-\bar p)\Gamma(p+q+1)}{\Gamma(-q-\bar p)
\Gamma(-q-\bar p -1)\gamma(-r)
\Gamma(p+1)\Gamma(p+q+r+2)}\quad .
\eeq

We now write the general result for the integral used in section
3 which was computed in reference \cite{fuku} (see also \cite{satoh} for
various equivalent expressions), namely
\ba
I & = & \int d^2u d^2v u^\alpha (1-u)^\beta\bar u^{\bar \alpha}
(1-\bar u)^{\bar \beta}v^{\alpha^\prime}(1-v)^{\beta^\prime}
\bar v^{\bar\alpha^{\prime}}(1-\bar v)^{\bar\beta^{\prime}}
|u-v|^{4\sigma}\nonumber\\
& = & -\frac 14\left ( 
C^{12}[\alpha_i,\alpha_i^\prime]
P^{12}[\bar\alpha_i,\bar\alpha_i^{\prime}]+C^{21}[\alpha_i,\alpha_i^\prime]
P^{21}[\bar\alpha_i,\bar\alpha_i^{\prime}]
\right )\quad ,
\ea
where
\ba
C^{ab}[\alpha_i,\alpha_i^\prime ] &=& \frac {
\Gamma(1+\alpha_a+\alpha_a^\prime
-k^\prime)\Gamma(1+\alpha_b+\alpha_b^\prime
-k^\prime)}{\Gamma(k^\prime-\alpha_c-\alpha_c^\prime)}
\nonumber\\ &&\qquad\qquad\times\;
 G\left [ 
\begin{array}{c}
\alpha_a^\prime+1,\alpha_b+1, k^\prime-\alpha_c-\alpha_c^\prime \\
\alpha_a^\prime-\alpha_c+1, \alpha_b-\alpha_c^\prime +1
\end{array}\right ]\quad .
\ea
Here
\beq
G\left [ \begin{array}{c}
a,b,c\\
e,f
\end{array}\right ] = \frac {\Gamma(a)\Gamma(b)\Gamma(c)}{\Gamma(e)\Gamma(f)}
{_3 F_2(a,b,c;e,f;1)} \quad ,
\eeq

\beq
\begin{array}{cccc}
\alpha_1=\alpha, &\alpha_2=\beta , &\alpha_3 = \gamma , &
\alpha + \beta + \gamma + 1 = k^\prime = -2\sigma - 1 ,\nonumber\\
\alpha_1^\prime = \alpha^\prime, &\alpha_2^\prime =\beta^\prime , &
\alpha_3^\prime = \gamma^\prime , &\alpha^\prime + \beta ^\prime + 
\gamma^\prime + 1 = k^\prime = -2\sigma - 1 ,
\end{array}
\eeq
and similarly for $\bar\alpha_i$ and $\bar\alpha_i^\prime$. $P^{12}$ and 
$P^{21}$ are given by
\beq
\left [
\begin{array}{c}
P^{12}\\
P^{21}
\end{array} 
\right ] = A_{\beta}\left [
\begin{array}{c}
C^{23}\\
C^{32}
\end{array} \right ] = A_\alpha^T\left [ 
\begin{array}{c}
C^{31}\\
C^{13}
\end{array}\right ],
\eeq
with
\beq
A_{\beta}=-4\left [
\begin{array}{cc}
s(\beta)s(\beta^\prime ) & -s(\beta)s(\beta^\prime - k^\prime)\\
-s(\beta^\prime )s(\beta - k^\prime ) & s(\beta)s(\beta^\prime )
\end{array}
\right ] ,
\eeq
and $s(x)=sin(\pi x)$.

The following identities among ${_3 F_2}(a,b,c;e,f;1)= F
\left [\begin{array}{c}
a,b,c\\
e,f
\end{array} \right ]$ or $ G\left [\begin{array}{c}
a,b,c\\
e,f
\end{array} \right ]$ have been used in section 3 to obtain the two point
function of spectral flowed states,
\ba
G\left [ \begin{array}{c}
a,b,c\\
e,f
\end{array}\right ] & = & 
\frac {\Gamma(b)\Gamma(c)}{\Gamma(e-a)\Gamma(f-a)}
G\left [ \begin{array}{c}
e-a,f-a,u\\
u+b,u+c
\end{array}\right ] \nonumber\\
& = &
\frac {\Gamma(b)\Gamma(c)\Gamma(u)}{\Gamma(f-a)\Gamma(e-b)\Gamma(e-c)}
G\left [ \begin{array}{c}
a,e-b,e-c\\
e,u+a
\end{array}\right ] \quad ,\nonumber\\
G\left [ \begin{array}{c}
a,b,c\\
e,f
\end{array}\right ] & = & 
\frac {s(e-b)s(f-b)}{s(a)s(c-b)}
G\left [ \begin{array}{c}
b,1+b-e,1+b-f\\
1+b-c,1+b-a
\end{array}\right ] \nonumber\\
&&\qquad +\frac {s(e-c)s(f-c)}{s(a)s(b-c)}G\left [ \begin{array}{c}
c,1+c-e,1+c-f\\
1+c-b,1+c-a
\end{array}\right ]\; , \nonumber\\
\ea
where $u=e+f-a-b-c$.

\section{Appendix C.
Three point function including three $w=1$ fields}

In this appendix we aim at computing a three point function involving three
$w=1$ fields following the procedure introduced in section 5. Namely
we want to compute
\beq
\left<\Phi^{w
=1,j_{1}}_{J_{1},{\bar J}_{1}}(x_{1},z_{1})\Phi_{J_{2},{\bar J}_{2}}^{w=1,
j_2}(x_{2},z_{2})
\Phi^{w=1,j_{3}}_{J_{3},{\bar J}_{3}}(x_{3},z_{3})\right> \; ,
\label{tre3}
\eeq
which should be obtained from the six point function including three
spectral flow operators
\beq
\langle
\Phi_{j_{1}}(y_{1},\zeta_{1})\Phi_{\frac
k2}(x_{1},z_{1})
\Phi_{j_{2}}(y_{2},\zeta_{2})
\Phi_{\frac k2}(x_{2},z_{2})\Phi_{\frac k2}(x_{3},z_{3})
\Phi_{j_{3}}(y_{3},\zeta_{3})\rangle\;
. \label{814}
\eeq
The fusion of the spectral flow operators with the remaining
fields is expected to give rise to the $w=1$ fields in (\ref{tre3}). 
Here we propose an approach which is alternative to the exhaustive one in 
section 3, and will 
allow us to compute (\ref{tre3}) for specific relations among the spins of 
the fields. In
fact, this will be done for the following cases

\ba
&i)& \quad j_{1}+j_{2}+j_{3}=\frac{k}{2}\; ,
\label{q1}\\
&ii)& \quad j_{1}+j_{2}-j_{3}=\frac{k-2}{2}\quad {\rm and\;
permutations}\; ,\label{q2}\\
&{\rm and}\nonumber\\
&iii)& \quad j_{1}+j_{2}-j_{3}=\frac{4-k}{2}\quad {\rm and\; permutations}
\; .
\label{q3}
\ea
The procedure exemplified here may be useful in order to compute 
particular expressions for correlators involving many units of spectral 
flow, where exhaustive calculations imply increasing difficulties.

Our strategy is as follows. We will first perform an intermediate
step in which we will spectral flow only two operators in (\ref{814}) in
order to
obtain the modified KZ and null vector equations to be satisfied by the
following four point function (after redefining
$(y_{3},\zeta_{3})\longrightarrow (x_{4},z_{4})$)
\beq
{\cal A}^{w}_{4} \equiv
\left<\Phi^{w
=1,j_{1}}_{J_1,{\bar J_1}}(x_{1},z_{1})\Phi_{J_{2},\bar
J_{2}}^{w=1, j_2}(x_{2},z_{2})\Phi_{\frac k2}(x_{3},z_{3})
\Phi_{j_{3}}(x_{4},z_{4})
\right> \; .
\label{treww}
\eeq
Then we will propose an appropriate ${\it ansatz}$ for the solution, and
finally spectral flow one last time in order to find (\ref{tre3}).

So, by performing calculations analogous to those in the previous
sections, we
find that the above four point function obeys the following modified KZ
equation which arises from the spectral flow operator
inserted at $(x_{3},z_{3})$ in (\ref{814})

\ba
&&\left(j_{1}-J_{1}+\frac{k}{2}-1\right)\frac{x_{31}}{z^{2}_{31}}\;
{\cal A}^{w}_{4}(J_{1}+1,J_{2})+
\left(j_{2}-J_{2}+\frac{k}{2}-1\right)\frac{x_{32}}{z^{2}_{32}}\;
{\cal A}^{w}_{4}(J_{1},J_{2}+1)
\nonumber\\ &=&
{\Bigg [}-\frac{\partial}{\partial
z_{3}}+\frac{1}{z_{31}}\left(x_{31}\frac{\partial}{\partial
x_{1}}-J_{1}\right)
+\frac{1}{z_{32}}\left(x_{32}\frac{\partial}{\partial
x_{2}}-J_{2}\right)\nonumber\\
&&\qquad\qquad\qquad\qquad\qquad\qquad\qquad\qquad\qquad +\;
\frac{1}{z_{43}}\left(x_{43}\frac{\partial}{\partial
x_{4}}+j_{3}\right)
{\Bigg ]}\; {\cal A}^{w}_{4}(J_{1},J_{2})\; .\nonumber\\
\label{800}
\ea
In addition, due to the three spectral flow operators in (\ref{814}) we
accordingly get three modified null vector equations for ${\cal
A}^{w}_{4}$. They read

\ba
&&\left(j_{1}-J_{1}+\frac{k}{2}-1\right)\frac{x^{2}_{31}}{z^{2}_{31}}\;
{\cal A}^{w}_{4}(J_{1}+1,J_{2})+
\left(j_{2}-J_{2}+\frac{k}{2}-1\right)\frac{x^{2}_{32}}{z^{2}_{32}}\;
{\cal A}^{w}_{4}(J_{1},J_{2}+1)
\nonumber\\ &=&
{\Bigg [}\frac{x_{31}}{z_{31}}\left(x_{31}\frac{\partial}{\partial
x_{1}}-2J_{1}\right)
+\frac{x_{32}}{z_{32}}\left(x_{32}\frac{\partial}{\partial
x_{2}}-2J_{2}\right)
-
\frac{x_{43}}{z_{43}}\left(x_{43}\frac{\partial}{\partial
x_{4}}+2j_{3}\right)
{\Bigg ]}\; {\cal A}^{w}_{4}(J_{1},J_{2})\; ,\nonumber\\
\label{801}
\ea

\ba
&&-\left(j_{1}+J_{1}-\frac{k}{2}-1\right)\;
{\cal A}^{w}_{4}(J_{1}-1,J_{2})-
\left(j_{2}-J_{2}+\frac{k}{2}-1\right)\frac{x^{2}_{21}}{z^{2}_{21}}\;
{\cal A}^{w}_{4}(J_{1},J_{2}+1)
\nonumber\\ &=&
{\Bigg [}\frac{x_{21}}{z_{21}}\left(x_{21}\frac{\partial}{\partial
x_{2}}+2J_{2}\right)
+\frac{x_{31}}{z_{31}}\left(x_{31}\frac{\partial}{\partial
x_{3}}+k\right)
+
\frac{x_{41}}{z_{41}}\left(x_{41}\frac{\partial}{\partial
x_{4}}+2j_{3}\right)
{\Bigg ]}\; {\cal A}^{w}_{4}(J_{1},J_{2})\; ,\nonumber\\
\label{802}
\ea
and
\ba
&&\left(j_{1}-J_{1}+\frac{k}{2}-1\right)\frac{x^{2}_{21}}{z^{2}_{21}}\;
{\cal A}^{w}_{4}(J_{1}+1,J_{2})+
\left(j_{2}+J_{2}-\frac{k}{2}-1\right)\;
{\cal A}^{w}_{4}(J_{1},J_{2}-1)
\nonumber\\ &=&
{\Bigg [}\frac{x_{21}}{z_{21}}\left(x_{21}\frac{\partial}{\partial
x_{1}}-2J_{1}\right)
-\frac{x_{32}}{z_{32}}\left(x_{32}\frac{\partial}{\partial
x_{3}}+k\right)
-
\frac{x_{42}}{z_{42}}\left(x_{42}\frac{\partial}{\partial
x_{4}}+2j_{3}\right)
{\Bigg ]}\; {\cal A}^{w}_{4}(J_{1},J_{2})\; .\nonumber\\
\label{803}
\ea

From the results of section 4, we expect that
${\cal A}^{w}_{4}$ had the same functional form as an unflowed four point
function, but with appropriate modified expressions for the spins and
conformal dimensions of the $w=1$ fields. So we look for a solution
of the following form
\ba
{\cal A}^{w}_{4}&=&{\cal D}_{1}(j_{1},J_{1}, j_{2},J_{2},j_{3})\;
{\cal D}_{2}(j_{1},{\bar J}_{1},j_{2},{\bar J}_{2},j_{3})
\;
{\cal F}(z,x)\; {\bar{\cal F}}({\bar z},{\bar x})\;
\nonumber\\&\times&
\left(x_{43}^{J_{1}+J_{2}-k/2-j_{3}}
x_{42}^{-2J_{2}}x_{41}^{J_{2}+k/2-J_{1}-j_{3}}
x_{31}^{j_{3}-J_{1}-J_{2}-k/2}\right)
\nonumber\\
&\times &
\left(z_{43}^{\Delta^{w=1}_{1}+\Delta^{w=1}_{2}+k/4-\Delta_{3}}
z_{42}^{-2\Delta^{w=1}_{2}}
z_{41}^{\Delta^{w=1}_{2}-k/4-\Delta^{w=1}_{1}-\Delta_{3}}
z_{31}^{\Delta_{3}-\Delta^{w=1}_{1}-\Delta^{w=1}_{2}+k/4}\right)
\nonumber\\ &\times & ({\rm antiholomorphic\; part})
\; .
\label{24}
\ea
Here we are assuming that there is only one state in each
intermediate channel, similarly as in the construction of
\cite{malda3}. We have shown in  Appendix A how the spins of the
intermediate states are fixed due to the presence of the spectral flow
operators, thus avoiding the integral introduced in \cite{tesch2}.
Notice that ${\cal D}_{1}$ and ${\cal D}_{2}$ are the
coefficients of the four point function depending on
the right and left spins of the string states.
The cross ratios $z, x$ are as in 
(\ref{l24}).

Now we make the following ${\it ansatz}$ for the functions ${\cal
F}$, ${\bar {\cal F}}$
\beq
{\cal F}=z^{\alpha}(1-z)^{\beta}x^{\mu}(1-x)^{\nu}(z-x)^{\rho}\; ,\qquad
{\bar {\cal
F}}={\bar z}^{{\bar
\alpha}}(1-{\bar
z})^{{\bar\beta}}{\bar
x}^{{\bar\mu}}(1-{\bar x})^{{\bar\nu}}({\bar z}-{\bar x})^{{\bar\rho}}\;
, \label{27}
\eeq
where the factors of $z, 1-z, x$ and $1-x$ are
suggested by the standard structure of singularities in conformal
field theory and string theory, namely those appearing at the boundary of
the moduli space
when two or more vertex operator
insertions collide on the worldsheet. The dependence on $z-x$ was found in
reference \cite{malda3} where it was shown to be required by monodromy
invariance of the four point
amplitude when the holomorphic and antiholomorphic parts
are combined. The singularity at $z=x$ was interpreted there as
due to instanton effects.

In principle, no further poles  arise in
presence of spectral flowed states, as noticed $e.g.$ from direct
inspection of (\ref{d1}). In
fact, plugging (\ref{24}) and  (\ref{27}) into (\ref{800})-(\ref{803}) we 
find
that the ${\it ansatz}$ (\ref{27}) is the
solution for relations (\ref{q1})-(\ref{q3}) among the spins.
Calculations involve a last step in which we spectral flow the field
$\Phi_{j_{3}}$ in (\ref{treww}) in order to get the three point function
(\ref{tre3}). This is done using the prescription (\ref{1}) and the
integration procedure is similar as in section 3. Since instead of
computing (\ref{treww})
we could also find the three point function (\ref{tre3}) starting from
correlators

\beq
{\cal A'}^{w}_{4} \equiv
\left<\Phi_{j_{1}}(y,\zeta)\Phi_{\frac k2}(x_{1},z_{1})\Phi^{w
=1,j_{2}}_{J_2,{\bar J_2}}(x_{2},z_{2})
\Phi^{w
=1,j_{3}}_{J_3,{\bar J_3}}(x_{3},z_{3})
\right> \; ,
\label{treww1}
\eeq
or

\beq
{\cal A''}^{w}_{4} \equiv
\left<\Phi^{w
=1,j_{1}}_{J_1,{\bar J_1}}(x_{1},z_{1})
\Phi_{j_{2}}(y,\zeta)\Phi_{\frac k2}(x_{2},z_{2})
\Phi^{w
=1,j_{3}}_{J_3,{\bar J_3}}(x_{3},z_{3})
\right> \; ,
\label{treww2}
\eeq
 we must also require that the final result does not
depend on the intermediate path we follow to compute it, which imposes
further restrictions.

Following all
the prescriptions above, using also the results in Appendix A
 and after 
performing some algebra, we arrive at
the
following three point functions
for spins related as in (\ref{q1})-(\ref{q3})

\ba
&&\left<\Phi^{w
=1,j_{1}}_{J_{1},{\bar J}_{1}}(x_{1},z_{1})\Phi_{J_{2},{\bar J}_{2}}^{w=1,
j_2}(x_{2},z_{2})
\Phi^{w=1,j_{3}}_{J_{3},{\bar J}_{3}}(x_{3},z_{3})\right>
\nonumber\\
&\sim &
B(j_{1})B(j_{2})B(j_{3})C\left(\frac{k}{2}-j_{1},\frac{k}{2}-j_{2}
,\frac{k}{2}-j_{3}\right)
\nonumber\\ &\times &
\pi\;\gamma(1-2j_{1})\;\frac{\Gamma(j_{1}+J_{1}-k/2)}
{\Gamma(1-j_{1}+J_{1}-k/2)}
\;\frac{\Gamma(j_{1}-{\bar J}_{1}+k/2)}{\Gamma(1-j_{1}-{\bar J}_{1}+k/2)}
\nonumber\\ &\times &
\pi\;\gamma(1-2j_{2})\;\frac{\Gamma(j_{2}+J_{2}-k/2)}
{\Gamma(1-j_{2}+J_{2}-k/2)}
\;\frac{\Gamma(j_{2}-{\bar J}_{2}+k/2)}{\Gamma(1-j_{2}-{\bar J}_{2}+k/2)}
\nonumber\\ &\times &
\pi\;\gamma(1-2j_{3})\;\frac{\Gamma(j_{3}+J_{3}-k/2)}
{\Gamma(1-j_{3}+J_{3}-k/2)}
\;\frac{\Gamma(j_{3}-{\bar J}_{3}+k/2)}{\Gamma(1-j_{3}-{\bar J}_{3}+k/2)}
\nonumber\\ &\times &
\left(x_{32}^{J_{1}-J_{2}-J_{3}}
x_{31}^{J_{2}-J_{1}-J_{3}}x_{21}^{J_{3}-J_{1}-J_{2}}\right)
\nonumber\\ &\times &
\left(z_{32}^{\Delta^{w=1}_{1}-\Delta^{w=1}_{2}-\Delta^{w=1}_{3}}
z_{31}^{\Delta^{w=1}_{2}-\Delta^{w=1}_{1}-\Delta^{w=1}_{3}}
z_{21}^{\Delta^{w=1}_{3}-\Delta^{w=1}_{1}-\Delta^{w=1}_{2}}\right)
\nonumber\\ &\times &  ({\rm antiholomorphic\; part})\; ,
\label{q20}
\ea
up to some $k$ dependent coefficient.

Explicit results for other relations among the spins could in principle be
obtained using a more involved ${\it ansatz}$ than the one in (\ref{27}),
possibly involving an hypergeometric function.


\begin{thebibliography}{99}
\bibitem{malda3} J. Maldacena and H. Ooguri, {\it Strings in $AdS_{3}$ and 
the 
SL(2,R) WZW Model. Part 3: Correlation Functions }, Phys. Rev. {\bf D65} 
(2002) 106006.
\bibitem{tesch1} J. Teschner, {\it On structure constants and fusion rules in 
the $SL(2,{\bf C})/SU(2)$ WZNW Model}, Nucl. Phys. {\bf B546} (1999) 390.
\bibitem{tesch2} J. Teschner, {\it Operator product expansion and 
factorization in the $H_{3}^{+}$-WZNW Model}, Nucl. Phys. {\bf B571} 
(2000) 555.
\bibitem{malda1} J. Maldacena and H. Ooguri, {\it Strings in $AdS_{3}$ and 
the 
SL(2,R) WZW Model. Part 1: The Spectrum}, J. Math. Phys. {\bf 42} (2001)
2929.
\bibitem{malda2} J. Maldacena, H. Ooguri and J. Son, {\it Strings in 
$AdS_{3}$ and 
the 
SL(2,R) WZW Model. Part 2: Euclidean black hole}, J. Math. Phys. {\bf 42} 
(2001) 2961.
\bibitem{zamo2} V. G. Knizhnik and A. B. Zamolodchikov, {\it Current 
algebra 
and Wess-Zumino Model in two dimensions}, Nucl. Phys. {\bf B247} (1984) 
83.
\bibitem{zamo1} A. B. Zamolodchikov and V. A. Fateev, {\it Operator 
algebra 
and 
correlation functions in the two-dimensional Wess-Zumino SU(2) 
$\times$ SU(2) 
Chiral Models}, Sov. J. Nucl. Phys. {\bf 43} (1986) 657.
\bibitem{nunez1} G. Giribet and C. N\'u\~nez, {\it Correlators in $AdS_{3}$ 
string theory}, JHEP {\bf 0106} (2001) 010.\\
D. Hofman and C. N\'u\~nez, {\it Free field realization of superstring 
theory on AdS$_3$}, JHEP {\bf 0407} (2004) 019.
\bibitem{zamo3} A. Zamolodchikov, unpublished notes.
\bibitem{geronimo} J. S. Geronimo and H. Navelet, {\it On certain two 
dimensional integrals that appear in conformal field theory},
 J. Math. Phys. {\bf 44} (2003) 2293.
\bibitem{fuku} T. Fukuda and K. Hosomichi, {\it Three-point functions in 
Sine-Liouville theory}, JHEP {\bf 0109} (2001) 003.
\bibitem{satoh} Y. Satoh, {\it Three point functions and operator product
expansion in the SL(2) conformal field theory}, Nucl. Phys. {\bf B629} 
(2002) 188.
\bibitem{ribault} S. Ribault, {\it Knizhnik-Zamolodchikov equations and 
spectral flow in $AdS_{3}$ string theory}, JHEP {\bf 0509} (2005) 045.
\bibitem{andreev} O. Andreev, {\it Operator algebra of the SL(2) conformal
field theories}, Phys. Lett. {\bf B363} (1995) 166.
\bibitem{giribet} G. Giribet, {\it On spectral flow symmetry and 
Knizhnik-Zamolodchikov equation}, Phys. Lett. {\bf B628} (2005) 148.
\end{thebibliography}
\end{document}